\documentclass[journal]{IEEEtran}
\usepackage{amsmath,graphicx}
\usepackage{cite}
\usepackage[caption=false,font=normalsize,labelfont=sf,textfont=sf]{subfig}
\usepackage{amsfonts,amsmath}
\DeclareMathOperator*{\argmax}{arg\,max}
\usepackage{multirow}
\usepackage{graphicx}

\usepackage{hyperref}
 \usepackage{pdfpages} 
\hypersetup{
    colorlinks=true,
    linkcolor=black,
    filecolor=black,      
    urlcolor=blue,
}
 \usepackage{algorithm}
\usepackage{algpseudocode}
\usepackage{breqn}
\usepackage{dblfloatfix}

\begin{document}


\title{\huge Intelligent and Reconfigurable Architecture for KL Divergence Based  Online Machine Learning Algorithm}

	\author{S. V. Sai Santosh and Sumit J. Darak
					\thanks{This work is supported by the funding received from DST INSPIRE faculty award and Core Research Grant from DST-SERB, India.}
			\thanks{S. V. Sai Santosh and Sumit J. Darak are with Electronics and Communications Department, 
				IIIT-Delhi, India-110020 (e-mail: \{siripurapu17197,sumit\}@iiitd.ac.in)}
}



	\maketitle
	\begin{abstract}
	
	Online machine learning (OML) algorithms do not need any training phase and can be deployed directly in an unknown environment. OML includes multi-armed bandit (MAB) algorithms that can identify the best arm among several arms by achieving a balance between exploration of all arms and exploitation of optimal arm. The Kullback-Leibler divergence based upper confidence bound (KLUCB) is the state-of-the-art MAB algorithm that optimizes exploration-exploitation trade-off but it is complex due to underlining optimization routine. This limits its usefulness for robotics and radio applications which demand integration of KLUCB with the PHY on the system on chip (SoC). In this paper, we efficiently map the KLUCB algorithm on SoC by realizing optimization routine via alternative synthesizable computation without compromising on the performance. The proposed architecture is dynamically reconfigurable such that the number of arms, as well as type of algorithm, can be changed on-the-fly. Specifically, after initial learning, on-the-fly switch to light-weight UCB offers around 10-factor improvement in latency and throughput. Since learning duration depends on the unknown arm statistics, we offer intelligence embedded in architecture to decide the switching instant. We validate the functional correctness and usefulness of the proposed architecture via a realistic wireless application and detailed complexity analysis demonstrates its feasibility in realizing intelligent radios.

	\end{abstract}
\vspace{-0.5cm}
	\section{Introduction}
Online machine learning (OML) algorithms such as multi-armed bandit (MAB) and reinforcement learning offer a simple but very powerful framework to enable decision making over time in an unknown uncertain environment \cite{MAB1, MAB2}. The MAB algorithm aims to identify the best arm among several arms and various extensions such as multi-play, multi-player, adversarial, contextual, and linear MABs have been explored to cater a wide range of applications \cite{MAB1}. Few of them include content (news or advt.) selection to maximize the number of clicks, dynamic pricing to maximize the total profit, medical trials to identify suitable drug and resource selection in robotics, data-center, IoT and wireless networks \cite{MAB1, MAB2, MAB3}. 

An optimal MAB algorithm guarantees logarithmic regret (i.e. loss due to sub-optimal arm selection) by achieving a balance between exploration of all arms to gain knowledge that may improve future performance and exploitation of an arm which can maximize the immediate performance.
The Kullback-Leibler divergence based upper confidence bound (KLUCB) is the state-of-the-art MAB algorithm that optimizes such trade-off but it is computationally complex due to underlining optimization routine \cite{klucb}. Other algorithms include UCB, Bayes\_UCB and Thompson Sampling (TS) which incur higher regret than KLUCB \cite{MAB1, MAB2, MAB3}. From an architecture perspective, none of these algorithms have ever been realized on the hardware. The usefulness of MAB algorithms in robotics, IoT and wireless applications and strict latency constraints demand efficient mapping to area and power-efficient architecture and tight integration with the physical layer (PHY) algorithms \cite{CA1}.

In this paper, we explore synthesizable computation to replace the optimization routine in KLUCB along with the hardware-software co-design approach to efficiently map the KLUCB algorithm on the Zynq system on chip (ZSoC) and validate its performance. To the best of our knowledge, this work is the first attempt towards the hardware realization and performance analysis of MAB algorithms.
Next, we explore dynamic partial reconfiguration (DPR) to realize reconfigurable architecture that allows on-the-fly configuration of a number of arms as well as the type of algorithm. We demonstrate around 10-factor improvement in latency and throughput by enabling on-the-fly switch from KLUCB to light-weight UCB after initial learning. Since learning duration depends on the unknown arm statistics, intelligence embedded in architecture offers the capability to optimize switching instant. We validate the functional correctness and usefulness of the proposed architecture via a realistic wireless application and detailed complexity analysis demonstrates its feasibility in realizing intelligent radios/robots. Please refer to \cite{Github} for additional supplementary tutorial and source codes.

\vspace{-0.25cm}
	\section{Synthesizable UCB and KLUCB Algorithms}
	
	
	In MAB setup, each experiment consists of $N, n\in\{1,2,..,N\}$ sequential slots with $K, k\in\{1,2,..K\}$ arms and the aim is to select the arm with highest reward as many times as possible. However, the reward distribution of the arms is unknown and needs to be learned. In this paper, we limit our discussion to Bernoulli reward distribution though proposed architectures can be tuned for Exponential and Poisson distributions as well. We consider single-play MAB where the algorithm can select only one arm in each slot. The arm selected in slot $n$ is denoted by, $I_{n}$ and $R_n$ denotes the reward received for the selected arm (i.e. only one feedback in each slot). Both UCB and KLUCB algorithms select each arm once in the beginning (i.e. first $K$ slots). Thereafter, in each subsequent time slot, quality factor (QF), $Q(k,n)$ is calculated for each arm.
	In UCB, the value of $Q_u(k,n)$ is given by \cite{MAB1},
	
	\begin{equation}
	\label{qf_ucb}
	    Q_u(k,n) = \frac{X(k,n)}{T(k,n)} + \sqrt{\frac{\alpha \log(n)}{T(k,n)}}
	\end{equation}
	where
	\begin{equation}
	\label{X}
	    X(k,n) = X(k,n-1) + R_{n-1} \cdot \textbf{1}_{\{I_{n-1}==k\}} \quad \forall k
	\end{equation}
	\begin{equation}
	\label{T}
	    T(k,n) = T(k,n-1) + \textbf{1}_{\{I_{n-1}==k\}} \quad \forall k
	\end{equation}
	where $\textbf{1}_{cond}$ is an indicator function and it is equal to 1 (or 0) if the condition, $cond$ is TRUE (or FALSE). The parameter, $\alpha$, is an exploration factor that can take any value between 0.5 and 2.  Based on calculated QFs, the arm with the highest QF is selected and it is denoted by, $I_n$ \cite{MAB1}.
	\begin{equation}
	\label{I}
	    I_n = \argmax_k Q_u(:,n)
	\end{equation}
	
	In the literature, various extensions of UCB such as UCB\_V and UCB\_T have been discussed and they slightly differ in terms of number of the arithmetic operations in QF calculation. We have realized all three UCB algorithms on ZSoC and compared their performance in Section V. However, due to limited space constraints, we limit the discussion to UCB and KLUCB since KLUCB QF calculation is significantly different and needs computationally complex optimization routine along with KL divergence, $d$, as shown below \cite{klucb}. 
 \begin{equation}
 \label{qkl}
     Q_{kl}(k,n) = \max \bigg\{q \in \left [ 0,1 \right ],d\bigg(\frac{X(k,n)}{T(k,n)},q\bigg) \leq Y(k,n)\bigg\}
 \end{equation}
 where
 \begin{equation*}
     Y(k,n)=\frac{\log n+c\log(\log n)}{T(k,n)}
     \vspace{-0.15cm}
 \end{equation*}
and
\begin{equation}
\vspace{-0.1cm}
\label{kl}
    d(p,q) = p\log\bigg(\frac{p}{q}\bigg) + (1-p) \log\bigg(\frac{1-p}{1-q}\bigg)
\end{equation}
	
	To realize Eq.~\ref{qkl}, we need KL divergence computation for large possible values of $q\in[0,1]$ which makes QF computation extremely expensive to realize in hardware. To overcome this limitation, we present an alternative heuristic approach shown in Algorithm 1 for Bernoulli reward distribution. The main idea is to dynamically and intelligently refine the range of $q$ based on comparison of the KL divergence between learned arm statistic, i.e. $\frac{X(k,n)}{T(k,n)}$ and expected arm statistics, $m_{id}$ (line 8) with the exploration factor, $S_2$ (line 9). Number of iterations of \textit{for} loop, i.e. parameter $\beta$, depends on $\Delta>0$ which is the minimum gap between statistics of any two arms and we set $\beta = \frac{1}{\Delta}$. We denote $\hat{\mu}(k,n)=\frac{X(k,n)}{T(k,n)}$ which is the learned mean of the reward distribution of $k^{th}$ arm till slot $n$ and $\mu(k)$ is its actual value which is unknown. Then, we have,
	\begin{equation}
	    	\Delta= \min_{i,j \in [N], i\neq j} |\mu(i)- \mu(j)|
	\end{equation}
	 We assume $\epsilon > \Delta$ and hence, $\beta = \frac{1}{\epsilon}$. 

		\begin{algorithm}[!h]
    \caption{Modified $Q_{kl}(k,n)$ Calculation in KLUCB}
			
			\begin{algorithmic}[1]
				
				\State Input: $X(k,n), T(k,n), n$
				\State Parameter: $\beta$
				\State Output: $Q(k,n)$
    \State $l_{id} = S_1 = X(k,n)/T(k,n)$
    \State $S_2 = [\log n+c\log(\log n)]/T(k,n) $
    \State $u_{id} = \min(1, S_1 + \sqrt{S_2/2})$
    \For{$i=1:1:\beta$}
    \State $m_{id} =(l_{id} + u_{id})/2$ 
    \If{$d(S_1,m_{id})>S_2$}
    \State $u_{id} = m_{id}$
    \Else
    \State $l_{id} = m_{id}$
    \EndIf
    \EndFor
    \State $Q(k,n)=u_{id}$
    			\end{algorithmic}
		\end{algorithm}

\section{Proposed Architecture}
We first discuss the KLUCB architecture details followed by a brief discussion on modifications needed for UCB realization. Each slot in the KLUCB algorithm consists of three tasks: 1)~Initialization (only for first $K$ slots) and parameter update based on the reward feedback, 2)~QF calculation for each arm, and 3)~Arm selection using calculated QF values. 
    
    \subsection{Initialization and Parameter Update Block}
    The initialization (INIT) and parameter update block is identical in both algorithms. At the beginning of a new experiment ($n=0$), the algorithm enters into the INIT phase and its duration is $K$ slots. The aim is to select each arm only once and in our architecture, this is accomplished using a pseudo-random sequence generator of length $K$. 
    
\begin{figure}[!h]
 	\vspace{-0.15cm}
 	\centering
 	\captionsetup{justification=centering}
 	\includegraphics[width=0.45\textwidth]{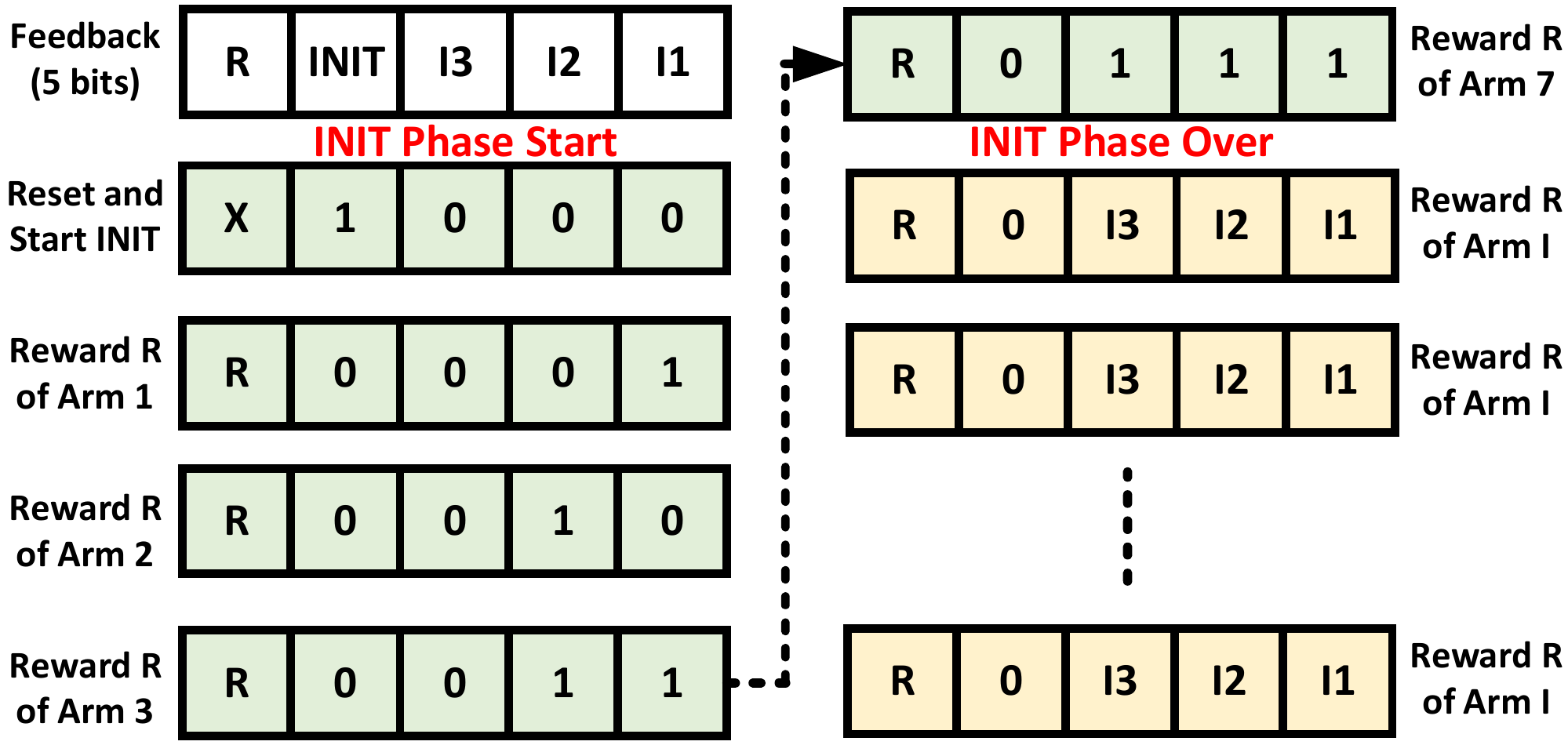}
 	\vspace{-0.25cm}
 	\caption{ Feedback signal format which is an input to parameter update block.}
 	\label{feedback}
 	\vspace{-0.15cm}
 \end{figure}

   In each slot, values of $\{X,T\}$ are updated based on the feedback as shown in Eq.~2 and 3 in Section II and the value of $n$ is incremented by 1.  The parameter update can be done at the end of the current slot after receiving the reward for the chosen arm or in the beginning of the subsequent slot. We choose the latter approach and hence, the feedback signal contains the information, $I_{n-1}$ and $R_{n-1}$ i.e., the reward received from the selected arm in $(n-1)^{th}$ slot. The format of the feedback signal is shown in Fig.~\ref{feedback}. The first bit is the reward (0 or 1 for Bernoulli case), second is the restart bit (1 to begin new experiment and 0 to continue the same experiment) and the remaining bits indicate the arm selected in the previous time slot resulting in total $\log_2(K_{max})+2$ bits. For easier understanding, the arms are shown to be selected in deterministic order in the INIT phase. For reward distributions with non-integer rewards, additional bits are needed.
 
\begin{figure}[!b]
 	\vspace{-0.15cm}
 	\centering
 	\captionsetup{justification=centering}
 	\includegraphics[width=0.45\textwidth]{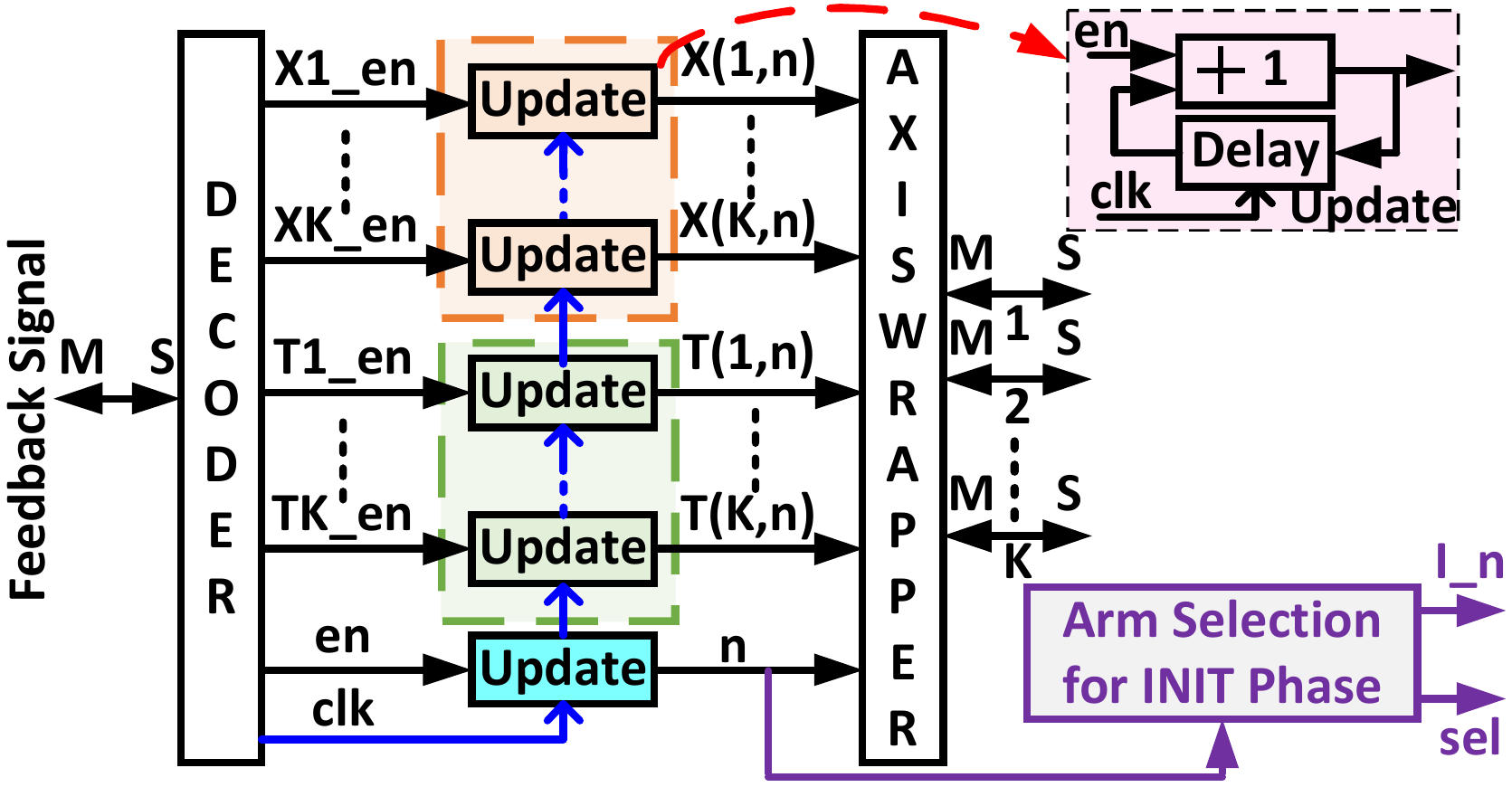}
 	\vspace{-0.25cm}
 	\caption{Architecture for initialization and parameter update block.}
 	\label{pua}
 	\vspace{-0.25cm}
 \end{figure}

    The architecture for the first task is shown in Fig.~\ref{pua}. In all figures, a double-headed arrow indicates AXI4 protocol where M and S symbols denote master and slave ports, respectively. In Fig.~\ref{pua}, the input decoder decodes the AXI4 feedback signal and generates various enable signals. For instance, $n\_{en}$ is generated once every slot which increments $n$ by 1 using \textit{update} block as shown in Fig.~\ref{pua}. If the $k^{th}$ arm is selected in the previous slot, then only $Tk\_{en}$ is generated. Similarly, $Xk\_{en}$ is generated only if $k^{th}$ arm is chose and its reward is 1. 

    \vspace{-0.2cm}
    
    \subsection{QF Calculation}

     	\begin{figure*}[!t]
 	\centering
 		\vspace{-0.15cm}
 	\captionsetup{justification=centering}
 	\includegraphics[width=0.85\textwidth]{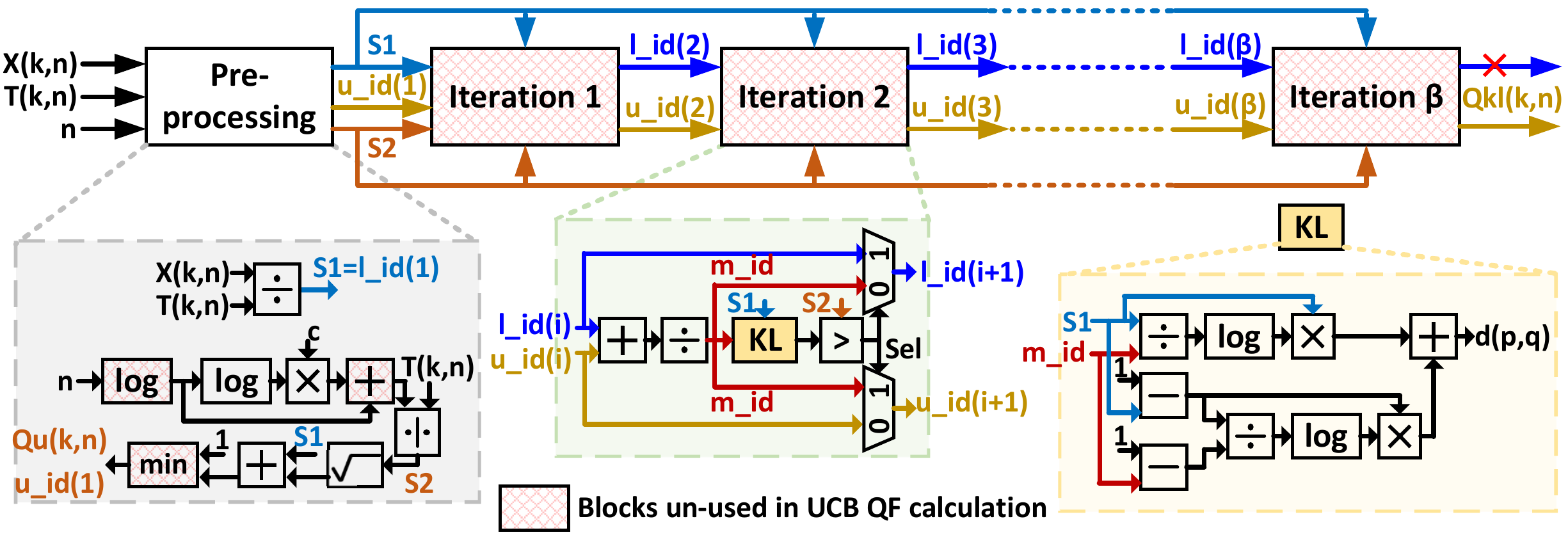}
 	\vspace{-0.25cm}
 	\caption{QF calculation for a single arm of KLUCB. Note that AXI dewrapper at the input, AXI wrapper at the output and AXI stream signals are omitted for maintaining the clarity of the figure.}
 	\label{qf}
 	\vspace{-0.45cm}
    \end{figure*}

 In each slot after the INIT phase, the algorithm calculates the value of QF for each arm using the updated parameters, $X(k,n), T(k,n)$ and $n$. We map the QF calculation steps of the modified KLUCB discussed in Algorithm 1 to the suitable architecture shown in Fig.~\ref{qf}. Since the QF calculation is identical for each arm and can be done in parallel without any interdependence, we limit our discussion to a single arm. 

The QF calculation needs three AXI4 stream inputs which are pre-processed to get $S_1, S_2, l_{id}$ and $u_{id}$ using the Steps 4-6 in Algorithm 1. Note that all operations are performed using the IPs with the AXI4 stream interface and for maintaining the clarify of architecture, we have omitted the AXI4 signals in Fig.~\ref{qf}. Furthermore, though reward, $X(k,n)$, can have only integer (1 and 0) values for Bernoulli distribution, the architecture supports single-precision floating-point arithmetic incurred in exponential and Poisson distributions. 

After pre-processing, QF calculation needs $\beta$ number of sequential loops. Note that due to interdependence between loops, these loops cannot be executed in parallel. For illustration, we have shown the architecture depicting various arithmetic operations and KL divergence (Eq.~\ref{kl}) calculation in each iteration of the loop. Necessary care has been taken to generate appropriate valid signals for each intermediate outputs so that the arithmetic blocks are enabled only when needed. Since only one iteration is active at a time, the same hardware is re-utilized for all $\beta$ iterations in KLUCB.

Using the reconfigurable and intelligent architecture discussed later in Section IV, the proposed architecture can switch to a light-weight UCB algorithm after the initial learning period. To highlight the need for such switch from a complexity perspective, the difference in computational complexity of QF calculations in two algorithms is highlighted in Fig.~\ref{qf}. It can be observed that QF UCB is obtained by disabling all iteration blocks of KLUCB along with logarithm, additional and minimum number identification sub-blocks in the pre-processing block (shown using the shaded pattern). Though savings in power and latency is evident, we need embedded intelligence and reconfigurability to auto-enable such switch.

    \vspace{-0.2cm}
     \subsection{Arm Selection}
     
\begin{figure}[!h]
 	\vspace{-0.25cm}
 	\centering
 	\captionsetup{justification=centering}
 	\includegraphics[width=0.425\textwidth]{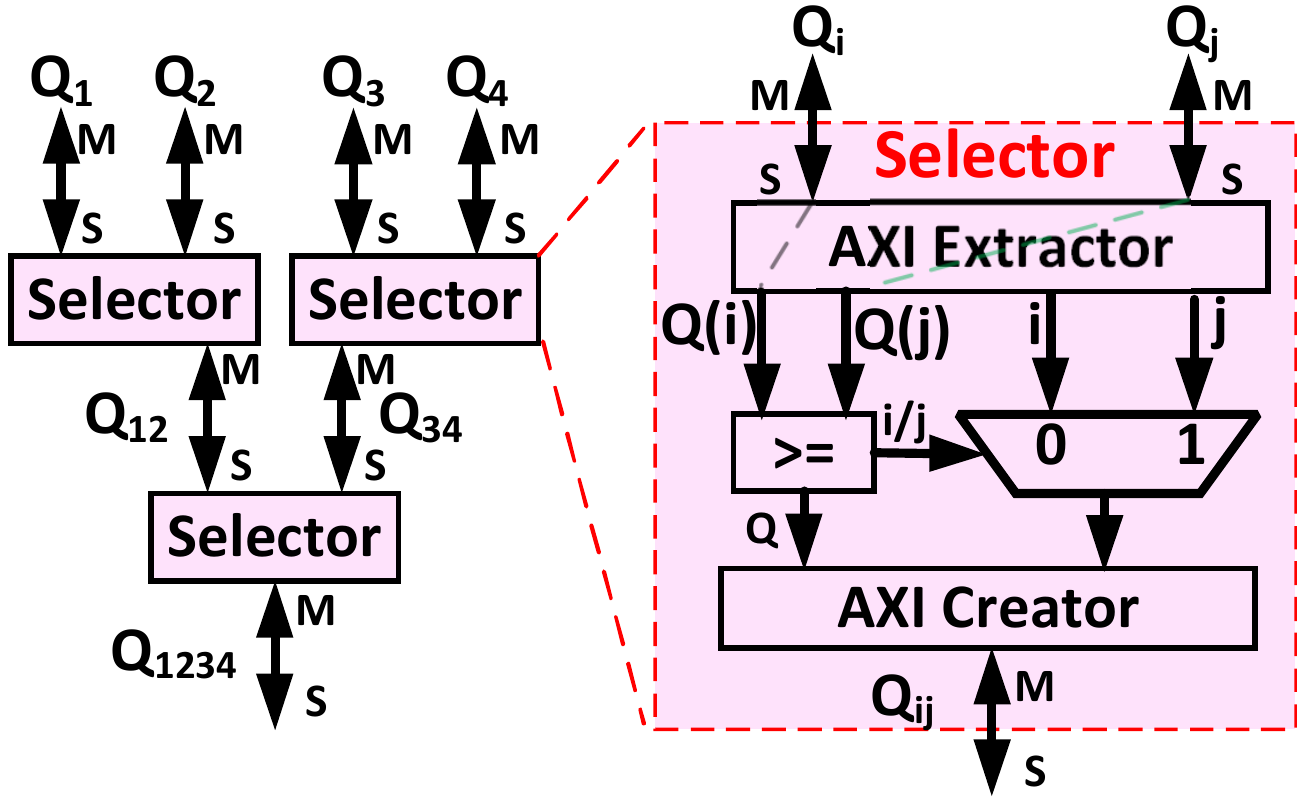}
 	\vspace{-0.25cm}
 	\caption{Arm selection block to select the arm with highest QF ($K=4$).}
 	\label{sel}
 	\vspace{-0.35cm}
 \end{figure}

    In each slot, a new arm with maximum QF value is selected by comparing the updated QF values as shown in Eq.~\ref{I}. Corresponding architecture for $K=4$ is shown in  Fig.~\ref{sel} where the output is $I_n$ i.e. index of an arm having the highest QF. Note that QF and arm selection blocks are bypassed in the INIT phase. Using these three blocks, proposed reconfigurable and intelligent architecture is presented in the next section.

	
	
 	\begin{table*}[!b]
\centering
\vspace{-0.4cm}
\caption{Comparison of Resource Utilization and Power Consumption}
\label{res}\vspace{-0.2cm}
\renewcommand{\arraystretch}{1.2}
\resizebox{\textwidth}{!}{%
\begin{tabular}{|c|c|c|c|c|c|c|c|c|c|c|c|c|c|}
\hline
\multirow{2}{*}{\textbf{Row}} & \multirow{2}{*}{\textbf{Architectures with $K_{max}$ = 4}} & \multicolumn{3}{l|}{\textbf{Slice LUT usage}} & \multicolumn{3}{l|}{\textbf{Slice DSP48 usage}} & \multicolumn{3}{l|}{\textbf{Slice FF usage}} & \multicolumn{3}{l|}{\textbf{Dynamic Power}} \\ \cline{3-14} 
                              &                                               & K=2           & K=3           & K=4           & K=2            & K=3            & K=4           & K=2           & K=3           & K=4          & K=2           & K=3          & K=4          \\ \hline
1                             & Reconfigurable Modified KLUCB                   & \textbf{23958}         & \textbf{37287}         & 50616         & \textbf{80}             & \textbf{120}            & \textbf{160}           & \textbf{6089}          & \textbf{10500}         & 14911        & \textbf{1.608}         & \textbf{1.647}        & \textbf{1.696}        \\ \hline
2                             & Conventional Modified KLUCB                               & 49980         & 49980         & \textbf{49980}         & 160            & 160            & \textbf{160}           & 14911         & 14911         & \textbf{14911}        & 1.697         & 1.697        & 1.697        \\ \hline \hline
3                           & Reconfigurable Modified KLUCB+UCB                   & \textbf{30175}         & \textbf{45122}         & 60069         & \textbf{92}             & \textbf{138}            & \textbf{184}           & \textbf{6718}          & \textbf{11233}         & \textbf{18748}        & \textbf{1.639}         & \textbf{1.723}        & \textbf{1.807}        \\ \hline

4                           & Conventional Modified KLUCB+UCB                     & 59433         & 59433         & \textbf{59433}         & 184             & 184            & \textbf{184}           & 18748          & 18748         & \textbf{18748}        & 1.807         & 1.807        & \textbf{1.807}        \\ \hline \hline

5                           & Reconfigurable Modified KLUCB+X                   & \textbf{31511}         & \textbf{47124}         & 62737         & \textbf{108}             & \textbf{162}            & \textbf{216}           & \textbf{10806}          & \textbf{15865}         & \textbf{20924}        & \textbf{1.648}         & \textbf{1.737}        & \textbf{1.826}        \\ \hline

6                           & Conventional Modified KLUCB+X                   & 62101         & 62101         & \textbf{62101}         & 216             & 216            & \textbf{216}           & 20924          & 20924         & \textbf{20924}        & 1.826         & 1.826        & \textbf{1.826}        \\ \hline

7                             & Only ARM based implementation                            & \multicolumn{9}{c|}{None. Only a single processor core used.}                                                                                  & 1.567         & 1.567        & 1.567        \\ \hline

\end{tabular}
}\vspace{-0.45cm}
\end{table*}

    \section{Intelligent and Reconfigurable Architecture}
    
       \begin{figure}[!b]
		\vspace{-0.2cm}
		\centering
		\captionsetup{justification=centering}
		\includegraphics[width=0.475\textwidth]{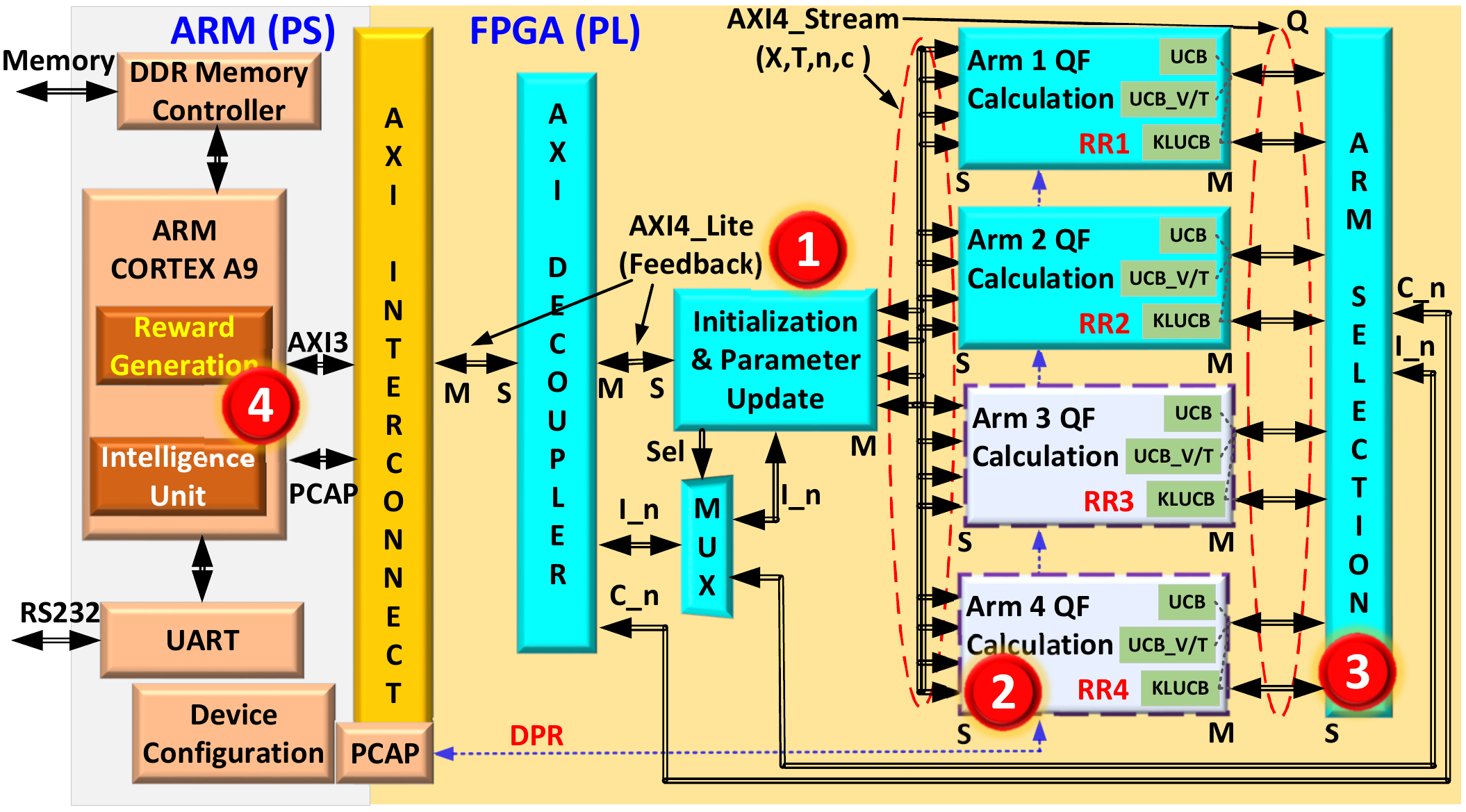}
		\vspace{-0.2cm}
		\caption{ Proposed DPR based intelligent and reconfigurable architecture for MAB algorithm with tunable number of arms and type of algorithm.}
		\label{block}
		\vspace{-0.2cm}
	\end{figure}
	
The proposed intelligent and reconfigurable architecture on ZSoC consisting of ARM processor (processing system i.e. PS) and FPGA (programmable logic i.e. PL) is shown in Fig.~\ref{block}. PL contains three blocks (1-3) corresponding to three tasks discussed in the previous section. The fourth task of generating the feedback signal with appropriate reward is realized in the ARM processor (PS) thereby making PS act as an environment. The INIT and parameter update block in PL can be integrated with the QF calculation block. However, resultant architecture demands four AXI4 handshakes between PS and PL in each slot which in turn incurs significantly penalty due to $K$ AXI write transactions compared to one transaction for architecture in Fig.~\ref{block}. Other realizations such as 1) Only ARM, and 2) ARM+NEON Co-processor, are also considered and please refer to Section V for details. 
	
The proposed architecture is made reconfigurable via the DPR property of ZSoC. Specifically, PS is responsible for generating appropriate DPR signals for changing the number of arms and types of algorithms, i.e., on-the-fly configuration to UCB, UCB\_T, UCB\_V and KLUCB algorithm. To enable such reconfiguration, we have incorporated PS controlled DPR via Processor Configuration Access Port (PCAP). For our architecture in Fig.~\ref{block} with $K_{max}=4$, we have four reconfiguration regions (RR), i.e. the region whose functionality can be changed on-the-fly. Since, each region can be configured with blank, UCB, UCB\_V, UCB\_T or KLUCB QF block, these five partial bit-streams are stored in the main memory or SD card. Via bare-metal application deployed on the ARM processor, the desired bit-streams are sent to the FPGA for appropriate RR configuration using the device configuration (DevC) direct memory access (DMA).
	
Reconfiguration of the number of arms depends on the environment and hence, can be user-controlled. Similarly, the user can decide the type of algorithm at the beginning of the experiment. The proposed architecture offers additional intelligence to automatically switch between the algorithms in an ongoing experiment to optimize latency and power without compromising on performance. For example, KLUCB is optimal because it reduces exploration by quick identification of optimal arm compared to UCB. This means though both KLUCB and UCB  are asymptotically optimal, i.e. they both can identify the optimal arm, KLUCB is better as it identifies the optimal arm in fewer exploration than UCB. Based on this observation, we embed additional intelligence in our architecture to deploy KLUCB in the initial slots and on-the-fly automatic switch to light-weight UCB after an initial learning period. As shown in Fig.~\ref{qf}, we obtain $Q_u(k,n)$ as well as $Q_{kl}(k,n)$ simultaneously. These two values are then compared in arm selector block to see whether both leads to the selection of the same arm. Based on this comparison, $C_n$ is generated which is 1 when the same arm is selected else 0. The intelligence unit in the ARM processor regularly checks $C_n$ over a suitably chosen window period and enable a switch to UCB if $C_n$ is observed 1 for the majority of times in the window period (indicating completion of the KLUCB exploration).

\begin{figure}[!h]
		\centering
		\captionsetup{justification=centering}
		\includegraphics[width=0.475\textwidth]{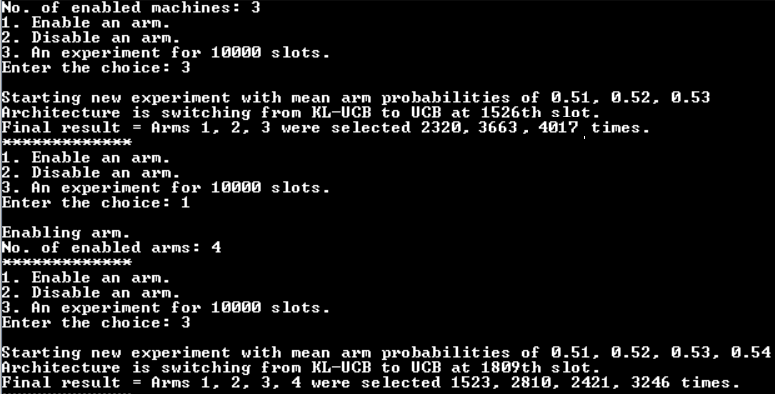}
		\vspace{-0.25cm}
		\caption{Snapshot of the baremetal application running on ARM depicting functionality of the proposed intelligent and reconfigurable architecture.}
		\vspace{-0.4cm}
		\label{scan}
	\end{figure}

In Fig.~\ref{scan}, we demonstrate the functioning of the proposed architecture. As shown, the user can add a new arm or remove the arm by choosing the appropriate option. When a user runs a new experiment with 3 machines, it can be observed that the switch between KLUCB to UCB happens at slot number 1526 and arm 3 with the highest reward is chosen maximum number of times. Next, the user adds new arm on-the-fly via DPR and in an experiment with four arms, algorithm needs more time for exploration which means KLUCB to UCB switch is delayed till slot 1809. As expected, arm 4 is chosen highest number of times. Please refer to \cite{Github,arxiv} containing a supplementary tutorial explaining detailed block design, DPR steps, and source codes.


	\section{Performance and Complexity Analysis}
	\label{result}
	
\begin{figure}[!h]
	 \vspace{-0.35cm}
     \centering
     \subfloat[]{\includegraphics[width=0.225\textwidth]{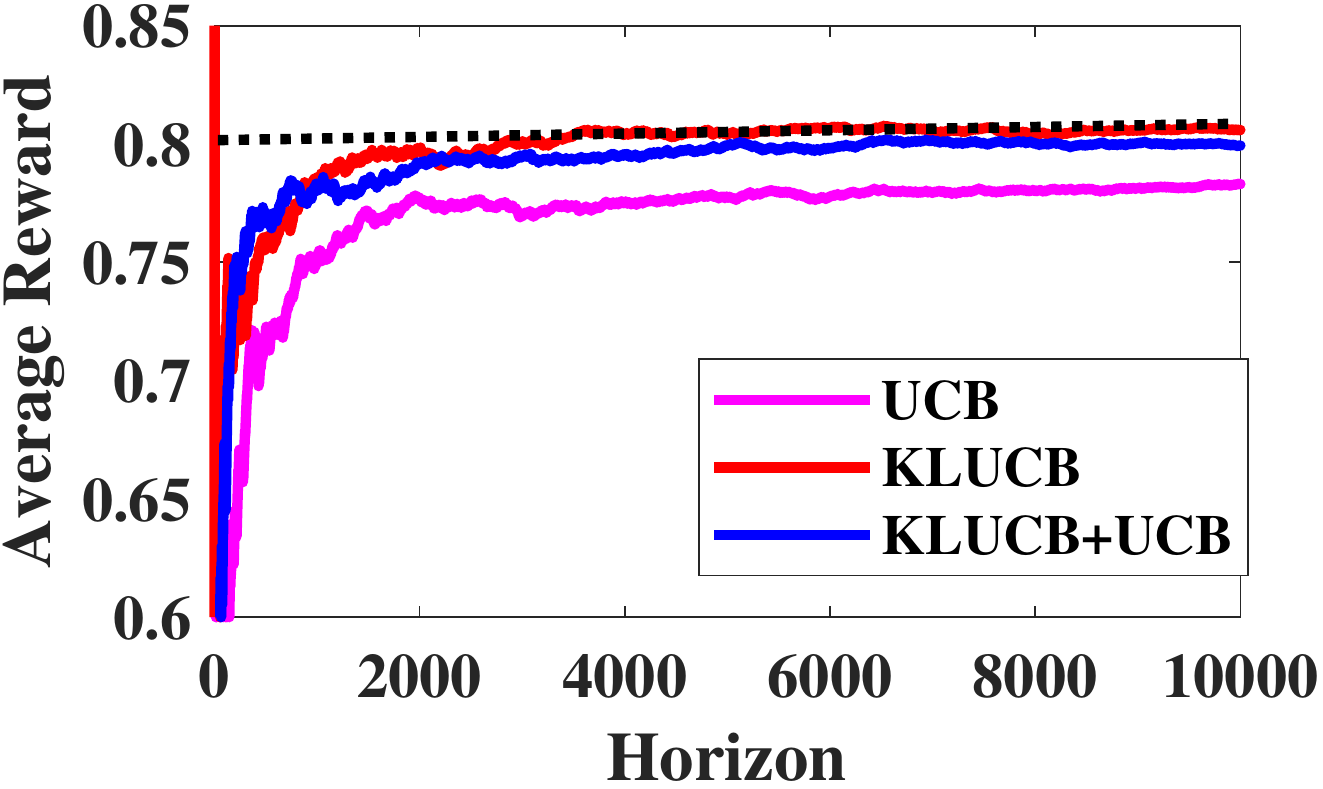}%
         \label{rew1}}
         \hspace{0.05cm}
     \subfloat[]{\includegraphics[width=0.225\textwidth]{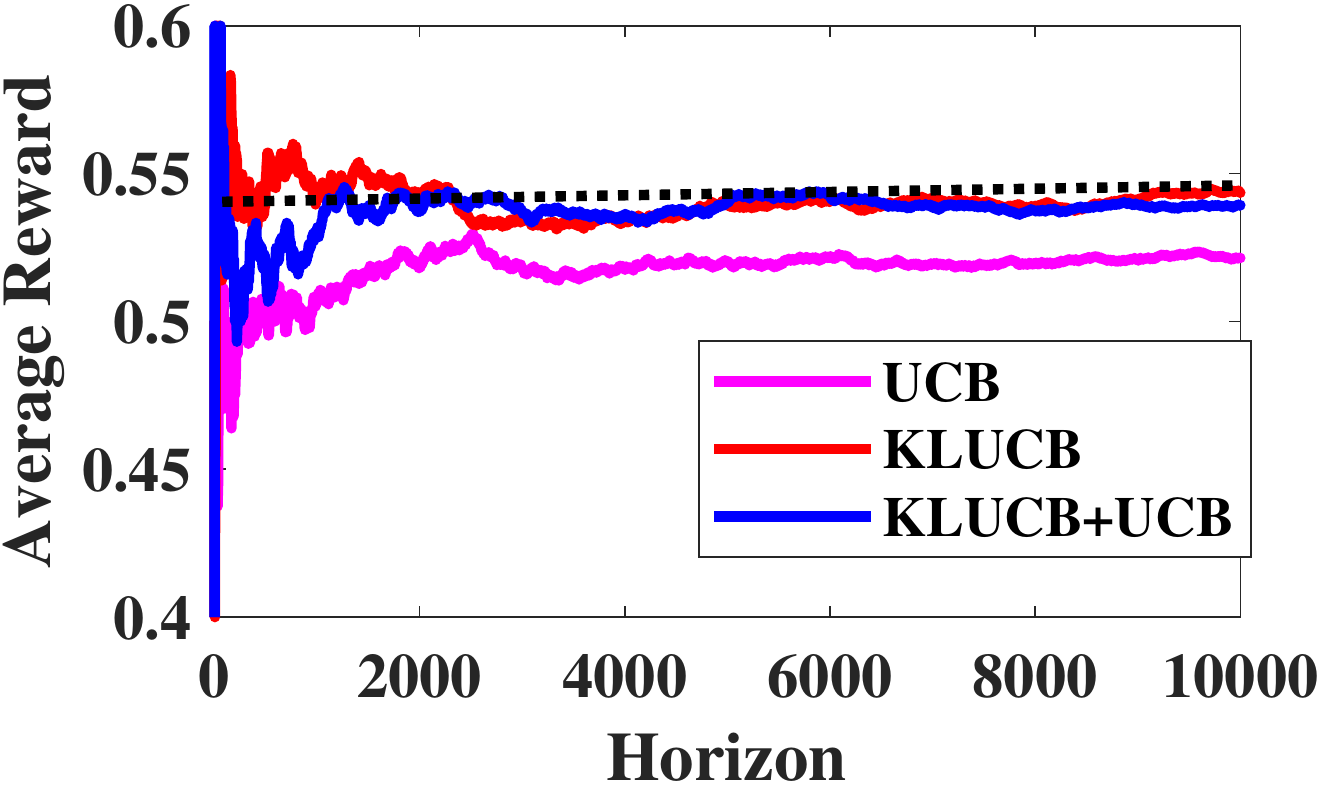}%
         \label{rew2}}
          \vspace{-0.15cm}
     \caption{ Average reward of UCB, optimization based KLUCB and integrated modified KLUCB (Algorithm 1) and UCB algorithm for (a) $\mu_1$, and (b) $\mu_2$. The horizontal dotted line indicates ideal reward.}
     \vspace{-0.15cm}
     \label{rew}
 \end{figure}

	To begin with, we compare the reward performance of the modified KLUCB in Algorithm 1 (referred to as KLUCB+UCB) with $\beta=16$, KLUCB \cite{klucb} and UCB \cite{MAB1} algorithms realized on ZSoC. We consider $K=4$ arms with a horizon consisting of $N=10000$ slots. The arms offer Bernoulli rewards with two different sets of mean distributions: 1)~$\mu_1=\{0.2, 0.4, 0.6, 0.8\}$ and $\mu_2=\{0.51, 0.52, 0.53, 0.54\}$. For easier analysis, the last arm has been chosen as the best arm i.e. arm with the highest reward. The average reward per slot for $\mu_1$ and $\mu_2$ is 0.54 and 0.8, respectively and this happens when the algorithm consistently selects the fourth arm. However, algorithms need exploration to learn arm distribution before converging to ideal average reward as shown in Fig.~\ref{rew}. As expected, optimization-based KLUCB offers the highest reward while proposed KLUCB+UCB (modified KLUCB with an intelligent switch to UCB after exploration) closely matches with the KLUCB and significantly outperforms UCB. 
	
	Next, we compare the resource utilization of various architectures in Table~\ref{res}. First, we consider two architectures for modified KLUCB discussed in Algorithm 1: 1) Reconfigurable via DPR, and 2) Velcro (conventional) approach. In the Velcro approach, all arms and hence, all QF calculation blocks are active at all times while the proposed reconfigurable architecture allows dynamic activation and deactivation of each arm. Note that none of these algorithms have been mapped to architectures yet in the literature. As shown in Row 1 and 2 of Table~\ref{res}, a reconfigurable approach offers lower resource utilization and power consumption except for a small increase in LUT when $K=K_{max}$. Similarly, we consider two more architectures for the proposed KLUCB+UCB algorithm where embedded intelligence enables the automatic switch from KLUCB to UCB. Corresponding results are given in Row 3 and 4 of Table~\ref{res}. In addition, two more architectures for the proposed KLUCB+X approach are considered where X stands for any one of the UCB extensions such as UCB, UCB\_V, and UCB\_T. Such architecture allows on-the-fly reconfiguration at the number of arms, type of algorithm as well as switching between the algorithms. 

All versions of the proposed reconfigurable architectures (Rows 2, 4, 6 in Table~\ref{res}) offer lower resource utilization and power consumption. Furthermore, when $K_{max}=4$ and $K=2$, the proposed approach consumes only 1.648W compared to 1.826W in the Velcro approach thereby offering around 5-10\% power saving for small-to-medium ($K/K_{max}$) ratio. Also, if we replace algorithm X in row 5 and 6 with more complex algorithms such as Bayesian UCB or TS or KLUCB extension or when the number of arms is large, i.e. $K_{max}>20$, there will be further improvement in these savings due to proposed reconfigurable approach. In Row 7, we consider only the ARM-based realization of the algorithm. Though its power consumption is around 1.567W and lowest due to hard processor, it has poor latency and hence, throughput as discussed next.

In Table~\ref{exetime}, we compare the execution time of various algorithms on three platforms: 1) Complex ZSoC (ARM + FPGA) as shown in Fig.~\ref{block}, 2) Only ARM, 3) ARM+NEON Co-processor. The execution time on the ARM is highest followed by the ARM+NEON platform while ZSoC offers the best performance validating the proposed hardware-software co-design approach. Between KLUCB and KLUCB+UCB approach, the latter offers more than 88\% reduction in execution time over the former. Thus, the proposed KLUCB+UCB approach offers lower execution time (Table~\ref{exetime}) without compromising on reward performance (Fig. 7). In wireless applications, MAB algorithms are realized in upper layers (MAC/Network) i.e. in ARM or other processors while the PHY is present in the SoC \cite{CA1,CA2}. The proposed architecture enables the shifting of the MAB algorithms from MAC to PHY layers along with an embedded intelligence unit that offers an accelerator factor ranging 50-100. Based on the results, we can say that the acceleration factor increases with the increase in $K$.
\begin{table}[!h]
\centering
\vspace{-0.25cm}
\caption{ Comparison of Execution Time Expressed in Microseconds ($\mu$s)}
\vspace{-0.25cm}
\label{exetime}
\renewcommand{\arraystretch}{1.2}
\resizebox{0.5\textwidth}{!}{%
\begin{tabular}{|c|c|c|c|c|c|c|c|}
\hline
\multirow{2}{*}{} & \multirow{2}{*}{\textbf{\begin{tabular}[c]{@{}c@{}}ZSoC\\ (in us)\end{tabular}}} & \multicolumn{3}{c|}{\textbf{\begin{tabular}[c]{@{}c@{}}PS (ARM)\\ (in us)\end{tabular}}} & \multicolumn{3}{c|}{\textbf{\begin{tabular}[c]{@{}c@{}}PS (ARM + NEON)\\ (in us)\end{tabular}}} \\ \cline{3-8} 
 &  & \textbf{K = 2} & \textbf{K = 3} & \textbf{K = 4} & \textbf{K = 2} & \textbf{K = 3} & \textbf{K = 4} \\ \hline
\textbf{UCB} & \textbf{792} & 7210 & 10820 & 15520 & 4170 & 6262 & 8349 \\ \hline
\textbf{UCB\_T} & \textbf{2773} & 18650 & 28970 & 37370 & 13109 & 19633 & 26187 \\ \hline
\textbf{UCB\_V} & \textbf{2760} & 17670 & 27290 & 34910 & 12673 & 19010 & 25361 \\ \hline
\textbf{KL\_UCB} & \textbf{30414} & 201458 & 302179 & 402900 & 164490 & 242163 & 331968 \\ \hline
\textbf{KL\_UCB + UCB} & \textbf{3756} & 26635 & 39956 & 54258 & 20202 & 29852 & 40711 \\ \hline
\textbf{KL\_UCB + UCB\_T} & \textbf{5531} & 46071 & 69951 & 92200 & 35816 & 53012 & 72054 \\ \hline
\textbf{KL\_UCB + UCB\_V} & \textbf{6078} & 39725 & 60277 & 79069 & 30891 & 45788 & 62154 \\ \hline
\end{tabular}%
}
\vspace{-0.25cm}
\end{table}

Next, we highlight the effect of $\beta$ on the performance of the KLUCB algorithm. As shown in Table~\ref{beta}, as the value of $\beta$ increases, the execution time increases due to sequential iterations in Algorithm 1. However, the rate of increases is substantially low in the proposed hardware-software co-design approach. In addition, we can see that the reward improves with $\beta$ and thus, the appropriate value of $\beta$ should be chosen to meet the desired trade-off between execution time and performance. In terms of learning performance, we observed that the error between actual and learned statistics decreases with an increase in $\beta$.
\begin{table}[!h]
\centering
\vspace{-0.25cm}
\caption{Execution Time vs. Reward Trade-off for Different $\beta$ Values}
\label{beta}
\renewcommand{\arraystretch}{1.2}
\resizebox{0.5\textwidth}{!}{%
\begin{tabular}{|c|c|c|l|c|l|}
\hline
\multirow{2}{*}{\textbf{}} & \multirow{2}{*}{\textbf{\begin{tabular}[c]{@{}c@{}}ZSoC\\ (Time in us, {[}$\mu_1$ Reward, $\mu_2$ Reward{]})\end{tabular}}} & \multicolumn{4}{c|}{\textbf{\begin{tabular}[c]{@{}c@{}}PS (ARM) \\ (Time in us, {[}$\mu_1$ Reward, $\mu_2$ Reward{]})\end{tabular}}} \\ \cline{3-6} 
 &  & \multicolumn{2}{c|}{\textbf{ARM + NEON}} & \multicolumn{2}{c|}{\textbf{Only ARM}} \\ \hline
\textbf{$\beta$ = 4} & \textbf{11700, {[}7718, 5330{]}} & \multicolumn{2}{c|}{101288, {[}7871, 5112{]}} & \multicolumn{2}{c|}{129592, {[}7701, 5294{]}} \\ \hline
\textbf{$\beta$ = 8} & \textbf{18100, {[}7774, 5112{]}} & \multicolumn{2}{c|}{169824, {[}7745, 5111{]}} & \multicolumn{2}{c|}{220122, {[}7771, 5281{]}} \\ \hline
\textbf{$\beta$ = 12} & \textbf{24500, {[}7856, 5411{]}} & \multicolumn{2}{c|}{242468, {[}7912, 5311{]}} & \multicolumn{2}{c|}{310345, {[}7832, 5304{]}} \\ \hline
\textbf{$\beta$ = 16} & \textbf{30414, {[}8102, 5387{]}} & \multicolumn{2}{c|}{331922, {[}8127, 5421{]}} & \multicolumn{2}{c|}{402900, {[}8106, 5430{]}} \\ \hline
\end{tabular}%
}
\vspace{-0.25cm}
\end{table}
	
	We also analyze the usefulness of the proposed architecture in for cognitive ad-hoc wireless networks where radio user aims to select the optimum channel for throughput maximization \cite{CA1}. Here, throughput refers to the number of bits transmitted per second (bps). We assume $K=4$, $N=10000$ and consider two different types of channels with statistics, $\mu$, randomly generated. As shown in Table~\ref{throughput}, proposed ZSoC based architecture offers a higher number of transmission of data bits as well as throughput than ARM+NEON based architecture. Another interesting observation is that KLUCB leads to the transmission of a higher number of bits but the proposed KLUCB+UCB offers significantly higher throughput due to lower execution time. For applications where FPGA is not available, the throughput of ARM+NEON based KLUCB+UCB realization is closed to that of ZSoC based KLUCB and at least 10 times higher than ARM-based KLUCB. We may be able to achieve further improvement in throughput if we select an appropriate UCB algorithm (UCB, UCB\_V or UCB\_T). Embedding such intelligence to select an algorithm is a challenging problem and focus of future work.

\begin{table}[!h]
\centering
\vspace{-0.25cm}
\caption{Throughput Comparison for Cognitive Ad-Hoc Networks}
\vspace{-0.25cm}
\renewcommand{\arraystretch}{1.2}
\label{throughput}
\resizebox{0.45\textwidth}{!}{%
\begin{tabular}{|c|c|c|c|c|c|c|c|c|c|}
\hline
\multicolumn{2}{|c|}{\multirow{2}{*}{\textbf{Algorithm}}} & \multicolumn{2}{c|}{\textbf{Data (Kbits)}} & \multicolumn{2}{c|}{\textbf{Throughput (Mbps)}} \\ \cline{3-6} 
\multicolumn{2}{|c|}{} & \textbf{$\mu_1$} &  \textbf{$\mu_2$} & \textbf{$\mu_1$} &\textbf{$\mu_2$} \\ \hline
\multirow{2}{*}{\textbf{KLUCB}} & \textbf{ZSoC} & \textbf{458.18} & \textbf{309.65} & 15.10 & 10.20 \\ \cline{2-6} 
 & \textbf{ARM+NEON} & \textbf{457.81} & \textbf{310.01} & 1.4 & 0.93 \\ \hline
\multirow{2}{*}{\textbf{KLUCB+UCB}} & \textbf{ZSoC} & 453.43 & 295.54 & \textbf{120.7} &\textbf{88.2} \\ \cline{2-6} 
 & \textbf{ARM+NEON} & 452.21 & 300.87 & 11.1 & 7.4 \\ \hline
\end{tabular}%
}\vspace{-0.25cm}
\end{table}
	
	\section{Conclusions and Future Directions}
A novel intelligent, reconfigurable, fast and computationally efficient architecture for Kullback–Leibler based Upper Confidence Bound (KLUCB) algorithm is presented in this paper. The performance analysis based on average reward, execution time, resource utilization and throughput highlights the advantages of the proposed architecture and its suitability for applications such as intelligent radio based wireless networks. Building on this state-of-the-art platform, future work will focus on open research problems such as intelligence to select the algorithm as well as optimal adaption strategy in dynamic and uncertain environment.

\pagebreak
	\newpage


\section*{Supplementary material}

# Tutorial: Reconfigurable and Low Complexity Architecture for UCB Algorithms

Authors: S. V. Sai Santosh and Dr. Sumit J. Darak, IIIT Delhi

{siripurapu17197, sumit}@iiitd.ac.in

## Introduction

In this lab, you will use Vivado IPI and Software Development Kit to create a reconfigurable peripheral using ARM Cortex-A9 processor system on Zynq. You will use Vivado IPI to create a top-level design, which includes the Zynq processor system as a sub-module. During the PR flow, you will define four Reconfigurable Partitions having three Reconfigurable Modules (UCB_V, UCB and UCB_tuned). You will create multiple Configurations and run the Partial Reconfiguration implementation flow to generate full and partial bitstreams. You will use ZC706 to verify the design in hardware using a SD card to initially configure the FPGA, and then partially reconfigure the device using the PCAP under user software control.

## Github Link: https://github.com/Sai-Santosh-99/PR_UCB

## Design Description

The purpose of this lab exercise is to implement a design that can be dynamically reconfigurable using PCAP resource and PS sub-system. The system consists of four peripheral (arms), having three unique function calculation capabilities (UCB_V, UCB and UCB_tuned). Then, the output (Quality-factor) of these machines goes into a Comparator IP which selects the machine with the largest Q-factor value. The user verifies the functionality using a user application. The dynamic modules are reconfigured using the PCAP resource available through Device Configuration block. The design is shown in Figure 1.

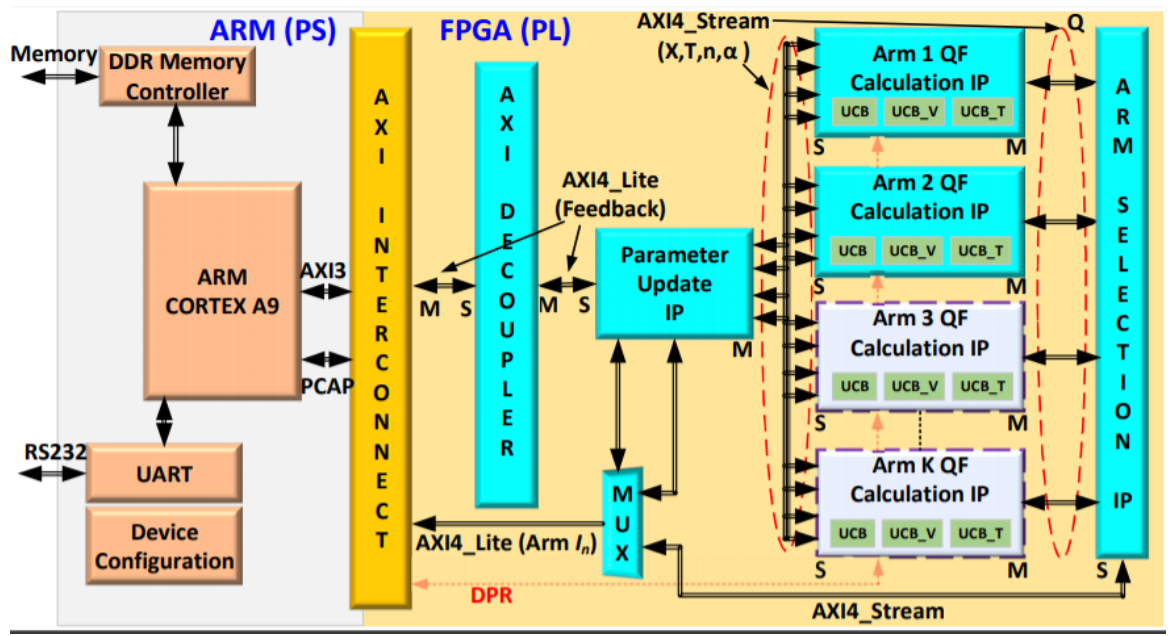

**Figure 1. The design**

The Sources directory provides the machine cores, source file for UCB_V, UCB and UCB_tuned, the software application (C code TestApp.c), and a place holder for the floorplan constraints (floorPlan.xdc). The Synth and its sub-directories structure will hold the synthesized checkpoints, the Implement and its sub-directories will hold the implemented configurations, the Checkpoint will hold the static, and the two configuration checkpoints, and the Bitstreams directory will hold the generated full and partial bitstreams. In the home directory, there are several Tcl scripts which will perform several tasks including the processor system creation and the bottom-up synthesis of the reconfigurable modules.

## Procedure

This lab is separated into steps that consist of general overview statements that provide information on the detailed instructions that follow. Follow these detailed instructions to progress through the lab.

## General Flow for this Lab

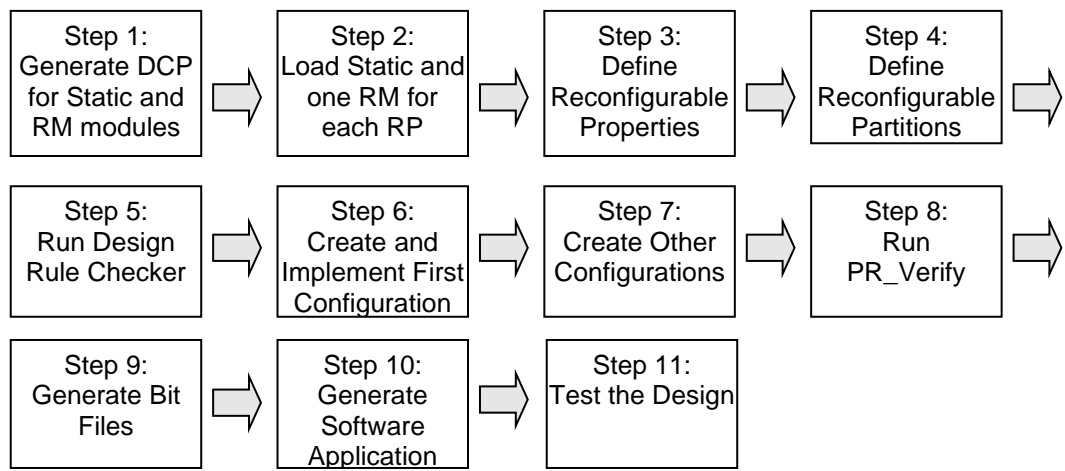

## Generate DCPs for the Static Design and RM Modules         Step 1

**1-1.** **Start Vivado and execute the provided Tcl script to create the design check point for the static design having one RP.**

**1-1-1.** Open **Vivado** by selecting **Start > All Programs > Xilinx Design Tools > Vivado 2018.2 > Vivado 2018.2**

**1-1-2.** In the Tcl Shell window enter the following command to change to the lab directory and hit **Enter**.

```
cd c:/Summer/Tutorial
```

**1-1-3.** Generate the PS design executing the provided Tcl script.

```
source blockDesign.tcl
```

This script will create the block design called system, instantiate ZYNQ PS with SD 0 and UART 1 interfaces enabled. It will also enable the GP0 interface along with FCLK0 and RESET0_N ports. The provided machines IP, PR decoupler and the Comparator will also be instantiated. It will then create a top-level wrapper file called system_wrapper.v which instantiates the system.bd (the block design).

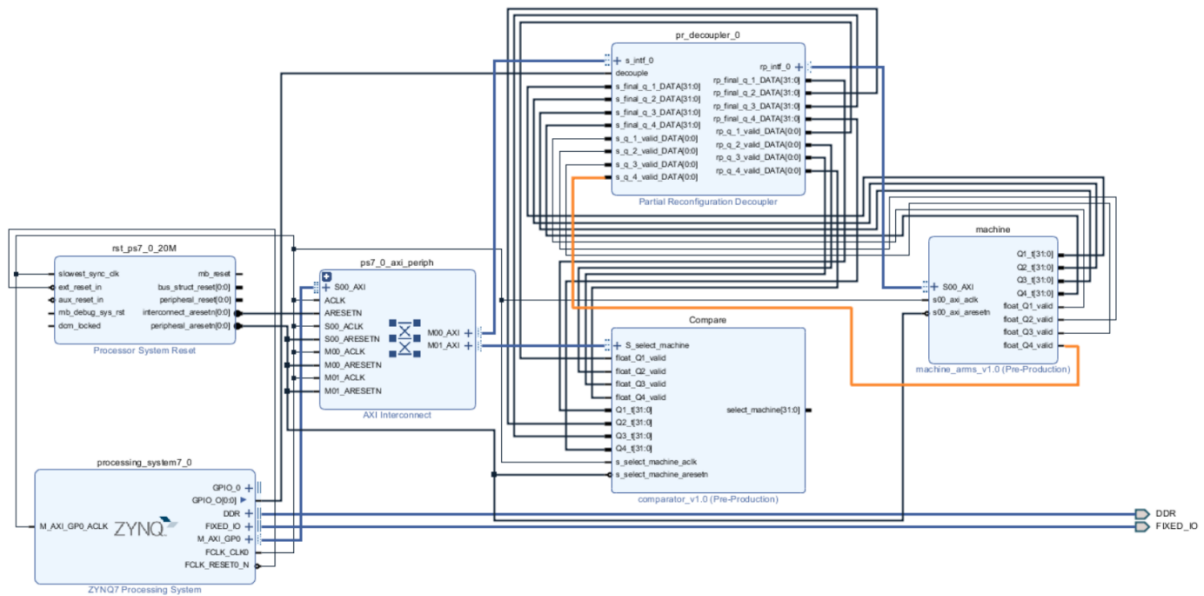

**Figure 2. The system block design**

**1-1-4.** Select the **Address Editor** tab. Expand **Data**. Expand **Unmapped Slaves**, if any, and right-click and select **Assign Address.**

**1-1-5.** Select **Tools** > **Validate Design.**

**1-1-6.** Select **File** > **Save Block Design.**

## 1-2. Synthesize the design to generate the dcp for the static logic of the design.

**1-2-1.** Click **Run Synthesis** under the *Synthesis* group in the *Flow Navigator* to run the synthesis process.

Wait for the synthesis to complete. When done click **Cancel**.

**1-2-2.** Using the windows explorer, copy the **system_wrapper.dcp** file from *tutorial\tutorial.runs\synth_*1 into the *Synth\Static* directory under the current lab directory.

**1-2-3.** Copy design checkpoints for the auto_pc, machine_arms, comparator, pr_decoupler_0, xbar_0, rst_ps7_0_20M, and processing_system7_0 instances to *Synth\Static* to sit alongside system_wrapper.dcp

**1-2-4.** Close the project by typing the `close_project` command in the Tcl console or selecting **File > Close Project**.

## 1-3. Since we have RMs in HDL format, we need to synthesize them and generate the dcp for each of the RMs. The generated dcps should be stored in appropriate directories so they can be accessed correctly; particularly, the dcp files for RM must be in separate directories as their dcp file names will be same for a given RP.

**1-3-1.** The HDL files for all the algorithms corresponding to each have been provided. Synthesize each each RM by creating a separate Vivado project. For the algorithm, synthesize the KL_UCB IP as kl_ucb, synthesize the UCB_V IP as q_variance, synthesize the UCB_T IP as q_tuned, and synthesize the UCB IP as Q_function. You can also see these IPs instantiated in the given HDL files.

**1-3-2.** Synthesize each of the RMs and write the design checkpoint (dcp) in the respective destination folder under the Synth directory. After each RM's dcp is generated, close the design.

**1-3-3.** At this point the directory content will look like shown below. Here, UCB_1 denotes that this dcp corresponds to the UCB algorithm for the first machine. You just need to synthesize each algorithm for each machine separately and add the corresponding .dcp files to the respective folders.

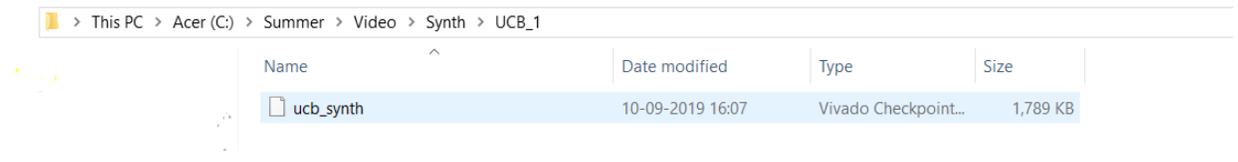

**Figure 7. Synth directory hierarchy and content**

## Load Static and one RM for the RP in Vivado                            Step 2

**Since all required netlist files (dcp) for the design are now available, you will use Vivado to floorplan the design, define Reconfigurable Partitions, add Reconfigurable Modules, run the implementation tools, and generate the full and partial bitstreams.**

**2-1.** **In this step you will load the static and one RM designs for the RP.**

**2-1-1.** In the Tcl Shell window enter the following command to change to the lab directory and hit **Enter**.

```
cd c:/Summer/Tutorial
```

**2-1-2.** Execute the following Tcl script to load the static design checkpoint.

```
source load_design_checkpoints.tcl
```

The script will do the following:

- Load the static design using the **open_checkpoint** command.

    ```
    open_checkpoint Synth/Static/system_wrapper.dcp
    ```

- Load the IP checkpoint for the Processing System by using the **read_checkpoint** command.

    ```
    read_checkpoint -cell system_i/processing_system7_0
    Synth/Static/system_processing_system7_0_0.dcp
    ```

- Load the IP checkpoint for the machines by using the **read_checkpoint** command.

```
read_checkpoint -cell system_i/machine
Synth/Static/system_machine_arms_0_0.dcp
```

- Load the IP checkpoint for the decoupler by using the **read_checkpoint** command.

   ```
   read_checkpoint -cell system_i/pr_decoupler_0
   Synth/Static/system_pr_decoupler_0_0.dcp
   ```

- Load the IP checkpoint for the Processing Reset by using the **read_checkpoint** command.

   ```
   read_checkpoint -cell system_i/rst_ps7_0_100M
   Synth/Static/system_rst_ps7_0_20M_0.dcp
   ```

- Load the IP checkpoint for the auto pc by using the **read_checkpoint** command.

   ```
   read_checkpoint -cell
   system_i/ps7_0_axi_periph/s00_couplers/auto_pc
   Synth/Static/system_auto_pc_0.dcp
   ```

- Load the IP checkpoint for the Comparator by using the **read_checkpoint** command.

   ```
   read_checkpoint -cell system_i/Compare
   Synth/Static/system_comparator_0_0.dcp
   ```

- Load the IP checkpoint for the system bus xbar by using the **read_checkpoint** command.

   ```
   read_checkpoint -cell system_i/ps7_0_axi_periph/xbar
   Synth/Static/system_xbar_0.dcp
   ```

**2-1-3.** Load one RM (UCB) for the RP by using the **read_checkpoint** command.

```
read_checkpoint -cell
system_i/machine/inst/machine_arms_v1_0_S00_AXI_inst/u1
Synth/UCB_1/ucb_synth.dcp

read_checkpoint -cell
system_i/machine/inst/machine_arms_v1_0_S00_AXI_inst/u2
Synth/UCB_2/ucb_synth.dcp

read_checkpoint -cell
system_i/machine/inst/machine_arms_v1_0_S00_AXI_inst/u3
Synth/UCB_3/ucb_synth.dcp

read_checkpoint -cell
system_i/machine/inst/machine_arms_v1_0_S00_AXI_inst/u4
Synth/UCB_4/ucb_synth.dcp
```

You can now see the design structure in the Netlist pane with an RM for the *u1, u2, u3 and u4* module loaded.

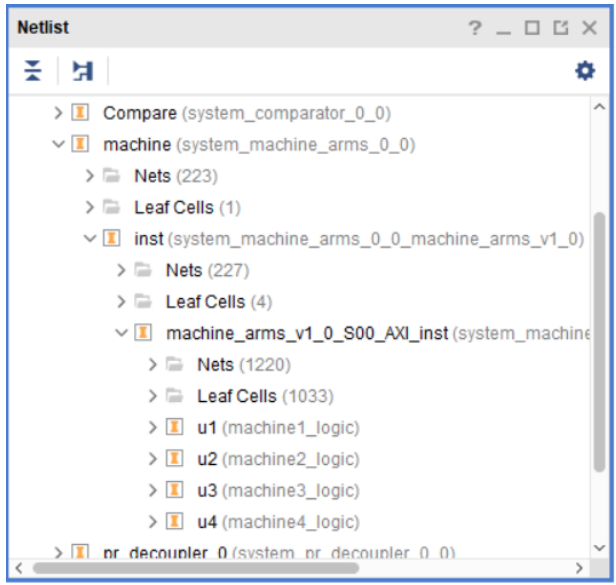

**Figure 9. Static design with RM loaded**

**2-1-4.** Select the *u1, u2 u3 and u4 instances one-by-one* and then select the *Properties* tab in the **Cell Properties** window. Note that the *IS_BLACKBOX* checkbox is not checked since a RM design is loaded.

## Define Reconfigurable Properties on each RM       Step 3

**3-1.  In this design you have one Reconfigurable Partition having two RMs. Define the reconfigurable properties to the loaded RM.**

**3-1-1.** Define each of the loaded RMs (submodules) as partially reconfigurable by setting the **HD.RECONFIGURABLE** property using the following commands.

```
set_property HD.RECONFIGURABLE 1 [get_cells
system_i/machine/inst/machine_arms_v1_0_S00_AXI_inst/u1]

set_property HD.RECONFIGURABLE 1 [get_cells
system_i/machine/inst/machine_arms_v1_0_S00_AXI_inst/u2]

set_property HD.RECONFIGURABLE 1 [get_cells
system_i/machine/inst/machine_arms_v1_0_S00_AXI_inst/u3]

set_property HD.RECONFIGURABLE 1 [get_cells
system_i/machine/inst/machine_arms_v1_0_S00_AXI_inst/u4]
```

This is the point at which the Partial Reconfiguration license is checked.

**3-1-2.** Select the `u1, u2, u3 and u4` instance and notice that the DONT_TOUCH checkbox is selected in the **Cell Properties** window.

## Define the Reconfigurable Partition Region

**3-2. Next you must floorplan the RP region. Depending on the type and amount of resources used by all the RMs for the given RP, the RP region must be appropriately defined so it can accommodate any RM variant.**

**3-2-1.** You execute the following command to define the region for each RP, perform the DRC.

```
read_xdc floorplan.xdc
```

## Create and Implement First Configuration

**4-1. Create and implement the first Configuration.**

**4-1-1.** Execute the following command to implement the first configuration, the UCB variant.

```
source create_first_configuration.tcl
```

The script will do the following tasks:

- The script will optimize, place and route the design by executing the following commands.

  ```
  opt_design
  
  place_design
  
  route_design
  ```

- Save the full design checkpoint.

  ```
  write_checkpoint -force Implement/UCB/top_route_design.dcp
  ```

  At this point, a fully implemented partial reconfiguration design from which full and partial bitstreams can be generated is ready. The static portion of this configuration **must** be used for all subsequent configurations, and to isolate the static design, the current reconfigurable module must be removed.

**4-2. After the first configuration is created, the static logic implementation will be reused for the rest of the configurations. So it should be saved. But before you save it, the loaded RM should be removed.**

**4-2-1.** Execute the following command to update the design with the blackbox and write the checkpoint.

```
source lock_placement_with_blackbox.tcl
```

The script will do the following tasks:

- Clear out the existing RMs executing the following commands.

  ```
  update_design -cell
  system_i/machine/inst/machine_arms_v1_0_S00_AXI_inst/u1 -
  black_box
  ```

```
update_design -cell
system_i/machine/inst/machine_arms_v1_0_S00_AXI_inst/u2 -
black_box

update_design -cell
system_i/machine/inst/machine_arms_v1_0_S00_AXI_inst/u3 -
black_box

update_design -cell
system_i/machine/inst/machine_arms_v1_0_S00_AXI_inst/u4 -
black_box
```

Issuing this command will result in design changes including, the number of Fully Routed nets (green) decreased, the number of Partially Routed nets (yellow) has increased, and *rp_instance* may appear in the Netlist view as empty.

- Lock down all placement and routing by executing the following command.

    ```
    lock_design -level routing
    ```

    Because no cell was identified in the `lock_design` command, the entire design in memory (currently consisting of the static design with black boxes) is affected.

- Write out the remaining static-only checkpoint by executing the following command.

    ```
    write_checkpoint -force Checkpoint/static_route_design.dcp
    ```

    This static-only checkpoint would be used for any future configuration, but here, you simply keep this design open in memory.

## Create Other Configurations

### 5-1. Read next set of RM dcp, create and implement the second configuration.

**5-1-1.** Execute the following command to create and implement the second configuration, the UCB_T variant.

```
source create_second_configuration.tcl
```

The script will do the following tasks:

- First, it will open the blanking configuration using the tcl command:

    ```
    open_checkpoint Checkpoint/static_route_design
    ```

- With the locked static design open in memory, read in post-synthesis checkpoint for the second reconfigurable module.

    ```
    read_checkpoint -cell
    system_i/machine1/inst/machine1_blank_v1_0_S00_AXI_inst/u1
    Synth/TUNED_1/ucb_synth.dcp
    ```

```
read_checkpoint -cell
system_i/machine2/inst/machine2_blank_v1_0_S00_AXI_inst/u1
Synth/TUNED_2/ucb_synth.dcp

read_checkpoint -cell
system_i/machine3/inst/machine3_blank_v1_0_S00_AXI_inst/u1
Synth/TUNED_3/ucb_synth.dcp

read_checkpoint -cell
system_i/machine4/inst/machine4_blank_v1_0_S00_AXI_inst/u1
Synth/TUNED_4/ucb_synth.dcp
```

Optimize, place and route the design by executing the following commands.

```
opt_design

place_design

route_design
```

- Save the full design checkpoint.

    ```
    write_checkpoint -force Implement/TUNED/top_route_design.dcp
    ```

- Close the project

    ```
    close_project
    ```

## Create Other Configurations

### 6-1. Read next set of RM dcp, create and implement the third configuration.

**6-1-1.** Execute the following command to create and implement the third configuration, the UCB_V variant.

```
source create_third_configuration.tcl
```

The script will do the following tasks:

- First, it will open the blanking configuration using the tcl command:

    ```
    open_checkpoint Checkpoint/static_route_design
    ```

- With the locked static design open in memory, read in post-synthesis checkpoint for the second reconfigurable module.

    ```
    read_checkpoint -cell
    system_i/machine1/inst/machine1_blank_v1_0_S00_AXI_inst/u1
    Synth/UCBV_1/ucb_synth.dcp
    ```

```
read_checkpoint -cell
system_i/machine2/inst/machine2_blank_v1_0_S00_AXI_inst/u1
Synth/UCBV_2/ucb_synth.dcp

read_checkpoint -cell
system_i/machine3/inst/machine3_blank_v1_0_S00_AXI_inst/u1
Synth/UCBV_3/ucb_synth.dcp

read_checkpoint -cell
system_i/machine4/inst/machine4_blank_v1_0_S00_AXI_inst/u1
Synth/UCBV_4/ucb_synth.dcp
```

Optimize, place and route the design by executing the following commands.

```
opt_design

place_design

route_design
```

- Save the full design checkpoint.

  ```
  write_checkpoint -force Implement/UCBV/top_route_design.dcp
  ```

- Close the project

  ```
  close_project
  ```

## Create Other Configurations

### 7-1. Read next set of RM dcp, create and implement the second configuration.

**7-1-1.** Execute the following command to create and implement the second configuration, the UCB_T variant.

```
source create_fourth_configuration.tcl
```

The script will do the following tasks:

- First, it will open the blanking configuration using the tcl command:

  ```
  open_checkpoint Checkpoint/static_route_design
  ```

- With the locked static design open in memory, read in post-synthesis checkpoint for the second reconfigurable module.

  ```
  read_checkpoint -cell
  system_i/machine1/inst/machine1_blank_v1_0_S00_AXI_inst/u1
  Synth/KL_1/ucb_synth.dcp
  ```

```
read_checkpoint -cell
system_i/machine2/inst/machine2_blank_v1_0_S00_AXI_inst/u1
Synth/KL_2/ucb_synth.dcp

read_checkpoint -cell
system_i/machine3/inst/machine3_blank_v1_0_S00_AXI_inst/u1
Synth/KL_3/ucb_synth.dcp

read_checkpoint -cell
system_i/machine4/inst/machine4_blank_v1_0_S00_AXI_inst/u1
Synth/KL_4/ucb_synth.dcp
```

Optimize, place and route the design by executing the following commands.

```
opt_design

place_design

route_design
```

- Save the full design checkpoint.

```
write_checkpoint -force Implement/KL/top_route_design.dcp
```

- Close the project

```
close_project
```

## 7-2. Create the blanking configuration.

**7-2-1.** Execute the following command to create and implement the second configuration

```
source create_blanking_configuration.tcl
```

The script will do the following tasks:

- Open the static route checkpoint.

```
open_checkpoint Checkpoint/static_route_design.dcp
```

- For creating the blanking configuration, use the `update_design -buffer_ports` command to insert LUTs tied to constants to ensure the outputs of the reconfigurable partition are not left floating.

```
update_design -buffer_ports -cell
system_i/machine/inst/machine_arms_v1_0_S00_AXI_inst/u1

update_design -buffer_ports -cell
system_i/machine/inst/machine_arms_v1_0_S00_AXI_inst/u2

update_design -buffer_ports -cell
system_i/machine/inst/machine_arms_v1_0_S00_AXI_inst/u3
```

```
                update_design -buffer_ports -cell
                system_i/machine/inst/machine_arms_v1_0_S00_AXI_inst/u4
```

- Now place and route the design.  There is no need to optimize the design.

    ```
    place_design

    route_design
    ```

    The base (or blanking) configuration bitstream, when we generate in the next section, will have no logic for either reconfigurable partition, simply outputs driven by ground.  Outputs can be tied to VCC if desired, using the HD.PARTPIN_TIEOFF property.

- Save the checkpoint in the BLANK directory.

    ```
    write_checkpoint -force Implement/BLANK/top_route_design.dcp
    ```

- Close the project

    ```
    Close_project
    ```

## Run PR_Verify

### 8-1. You must ensure that the static implementation, including interfaces to reconfigurable regions, is consistent across all Configurations.  To verify this, you run the PR_Verify utility

**8-1-1.** Run the **pr_verify** command from the Tcl Console.

```
source verify_configurations.tcl
```

The script will perform the following tasks:

- Execute the pr_verify command and then close the project:

    ```
    pr_verify -initial Implement/TUNED/top_route_design.dcp -
    additional {Implement/BLANK/top_route_design.dcp
    Implement/UCB/top_route_design.dcp
    Implement/UCBV/top_route_design.dcp}
    ```

You should see the message indicating the TUNED configuration is compatible with BLANK, and the TUNED configuration is compatible with UCB  & TUNED configuration is compatible with UCBV.

- Execute the following command to close the project.

    ```
    close_project
    ```

# Generate Bit Files

## 9-1. After all the Configurations have been validated by PR_Verify, full and partial bit files must be generated for the entire project

**9-1-1.** Generate the full configurations and partial bitstreams by executing the following tcl script.

```
source generate_bitstreams.tcl
```

**9-1-2.** The script will do the following tasks:

- Read the first configuration in the memory, using the command:

    ```
    open_checkpoint Implement/UCB/top_route_design.dcp
    ```

- Generate the full and partial bitstreams for this design.

    ```
    write_bitstream -file Bitstreams/UCB.bit

    write_cfgmem -format BIN -interface SMAPx32 -disablebitswap -loadbit "up 0 Bitstreams/UCB_pblock_machine1_partial.bit" Bitstreams/UCB_1.bin

    write_cfgmem -format BIN -interface SMAPx32 -disablebitswap -loadbit "up 0 Bitstreams/UCB_pblock_machine2_partial.bit" Bitstreams/UCB_2.bin

    write_cfgmem -format BIN -interface SMAPx32 -disablebitswap -loadbit "up 0 Bitstreams/UCB_pblock_machine3_partial.bit" Bitstreams/UCB_3.bin

    write_cfgmem -format BIN -interface SMAPx32 -disablebitswap -loadbit "up 0 Bitstreams/UCB_pblock_machine4_partial.bit" Bitstreams/UCB_4.bin

    close_project
    ```

    Notice the five bitstreams will be created.

- Generate full and partial bitstreams for the second configuration.

    ```
    write_cfgmem -format BIN -interface SMAPx32 -disablebitswap -loadbit "up 0 Bitstreams/UCBV_pblock_machine1_partial.bit" Bitstreams/UCBV_1.bin

    write_cfgmem -format BIN -interface SMAPx32 -disablebitswap -loadbit "up 0 Bitstreams/UCBV_pblock_machine2_partial.bit" Bitstreams/UCBV_2.bin

    write_cfgmem -format BIN -interface SMAPx32 -disablebitswap -loadbit "up 0 Bitstreams/UCBV_pblock_machine3_partial.bit" Bitstreams/UCBV_3.bin

    write_cfgmem -format BIN -interface SMAPx32 -disablebitswap -loadbit "up 0 Bitstreams/UCBV_pblock_machine4_partial.bit" Bitstreams/UCBV_4.bin
    ```

```
close_project
```

The five files will be created.

- Generate a full bitstream with black boxes, plus blanking bitstreams for the reconfigurable modules. Blanking bitstreams can be used to "erase" an existing configuration to reduce power consumption.

  ```
  write_bitstream -file Bitstreams/BLANK.bit
  ```

  ```
  write_cfgmem -format BIN -interface SMAPx32 -disablebitswap -
  loadbit "up 0 Bitstreams/BLANK_pblock_machine1_partial.bit"
  Bitstreams/BLANK_1.bin
  ```

  ```
  write_cfgmem -format BIN -interface SMAPx32 -disablebitswap -
  loadbit "up 0 Bitstreams/BLANK_pblock_machine2_partial.bit"
  Bitstreams/BLANK_2.bin
  ```

  ```
  write_cfgmem -format BIN -interface SMAPx32 -disablebitswap -
  loadbit "up 0 Bitstreams/BLANK_pblock_machine3_partial.bit"
  Bitstreams/BLANK_3.bin
  ```

  ```
  write_cfgmem -format BIN -interface SMAPx32 -disablebitswap -
  loadbit "up 0 Bitstreams/BLANK_pblock_machine4_partial.bit"
  Bitstreams/BLANK_4.bin
  ```

  ```
  close_project
  ```

- Generate full and partial bitstreams for the third configuration.

  ```
  write_bitstream -file Bitstreams/TUNED.bit
  ```

  ```
  write_cfgmem -format BIN -interface SMAPx32 -disablebitswap -
  loadbit "up 0 Bitstreams/TUNED_pblock_machine1_partial.bit"
  Bitstreams/TUNED_1.bin
  ```

  ```
  write_cfgmem -format BIN -interface SMAPx32 -disablebitswap -
  loadbit "up 0 Bitstreams/TUNED_pblock_machine2_partial.bit"
  Bitstreams/TUNED_2.bin
  ```

  ```
  write_cfgmem -format BIN -interface SMAPx32 -disablebitswap -
  loadbit "up 0 Bitstreams/TUNED_pblock_machine3_partial.bit"
  Bitstreams/TUNED_3.bin
  ```

  ```
  write_cfgmem -format BIN -interface SMAPx32 -disablebitswap -
  loadbit "up 0 Bitstreams/TUNED_pblock_machine4_partial.bit"
  Bitstreams/TUNED_4.bin
  ```

  ```
  close_project
  ```

  The five files will be created.

# About the C code

- This function is used for moving the partial binaries from the SD Card to the DDR memory.

```c
int SD_TransferPartial(char *FileName, u32 DestinationAddress, u32 ByteLength)
{
    FIL fil;
    FRESULT rc;
    UINT br;

    rc = f_open(&fil, FileName, FA_READ);
    if (rc) {
        xil_printf(" ERROR : f_open returned %d\r\n", rc);
        return XST_FAILURE;
    }

    rc = f_lseek(&fil, 0);
    if (rc) {
        xil_printf(" ERROR : f_lseek returned %d\r\n", rc);
        return XST_FAILURE;
    }

    rc = f_read(&fil, (void*) DestinationAddress, ByteLength, &br);
    if (rc) {
        xil_printf(" ERROR : f_read returned %d\r\n", rc);
        return XST_FAILURE;
    }

    rc = f_close(&fil);
    if (rc) {
        xil_printf(" ERROR : f_close returned %d\r\n", rc);
        return XST_FAILURE;
    }

    return XST_SUCCESS;
}
```

- This function is used to initialize the SD controller.

```c
int SD_Init()
{
    FRESULT rc;

    rc = f_mount(&fatfs, "", 0);
    if (rc) {
        xil_printf(" ERROR : f_mount returned %d\r\n", rc);
        return XST_FAILURE;
    }

    return XST_SUCCESS;
}
```

- Below, we have defined the function to write the data to our arms.

```
#define mWriteReg(BaseAddress, RegOffset, Data) \
    Xil_Out32((BaseAddress) + (RegOffset), (u32)(Data))
```

- Below, we have defined the function to read the data from the comparator.

```
#define mReadReg(BaseAddress, RegOffset) \
    Xil_In32((BaseAddress) + (RegOffset))
```

- Below is the pseudocode for the Multi-Armed Bandit algorithm on FPGA.

```
algo()
{
    for n times {

        •   Generate a random data value.

        •   Read the output from the Comparator

        •   If Comparator says Machine i is better-

            1. CHECK THE REWARD.

            2. INCREMENT X(i) if REWARD=1, else don't.

            3. INCREMENT T(i) as the Machine i was selected.

            4. WRITE THE CORRESPONDINT DATA TO THE FPGA CONVEYING THE
               MACHINE SELECTED AND ITS REWARD.
    }
}
```

## Generate the Software Application                                     Step 10

**10-1. Open the PS design that was created in Step 1. Export the hardware design and launch SDK.**

**10-1-1.** Click on the **Open Project** link, browse to *c:/Summer/Tutorial/tutorial*, select the *tutorial.xpr* and click **OK** to open the design created in Step 1.

**10-1-2.** Select **File > Export > Export Hardware…**

**10-1-3.** In the *Export Hardware* form, <u>do not check</u> the *Include bitstream* checkbox and click **OK**.

**10-1-4.** Select **File > Launch SDK**

**10-1-5.** Click **OK** to launch SDK.

The SDK program will open. Close the Welcome tab if it opens.

## 10-2. Create a Board Support Package enabling generic FAT file system library.

**10-2-1.** In **SDK**, select **File** > **New** > **Board Support Package.**

**10-2-2.** Click **Finish** with the default settings (with standalone operating system).

This will open the Software Platform Settings form showing the OS and libraries selections.

**10-2-3.** Select **xilffs** as the FAT file support is necessary to read the partial bit files.

| Name | Version | Description |
|---|---|---|
| libmetal | 1.0 | Libmetal Library |
| lwip141 | 1.6 | lwIP TCP/IP Stack library: lwIP v1.4.1 |
| openamp | 1.1 | OpenAmp Library |
| ☑ xilffs | 3.4 | Generic Fat File System Library |
| xilflash | 4.2 | Xilinx Flash library for Intel/AMD CFI com... |
| xilisf | 5.7 | Xilinx In-system and Serial Flash Library |
| xilmfs | 2.1 | Xilinx Memory File System |
| xilpm | 2.0 | Power Management API Library for Zynq... |
| xilrsa | 1.2 | Xilinx RSA Library |
| xilskey | 6.0 | Xilinx Secure Key Library |

**Figure 13. Selecting the xilffs library support**

**10-2-4.** Click **OK** to accept the settings and create the BSP.

## 10-3. Create an application.

**10-3-1.** Select **File** > **New** > **Application Project.**

**10-3-2.** Enter **TestApp** as the *Project Name*, and for *Board Support Package*, choose **Use Existing** (*standalone_bsp_0* should be the only option).

**10-3-3.** Click **Next**, and select *Empty Application* and click **Finish.**

**10-3-4.** Expand the **TestApp** entry in the project view, right-click the *src* folder, and select **Import.**

**10-3-5.** Expand **General** category and double-click on **File System.**

**10-3-6.** Browse to *c:\Summer\Tutorial\Sources* and click **OK.**

**10-3-7.** Select **TestApp.c** and click **Finish** to add the file to the project.

**10-3-8.** Right-click on **TestApp** and select **C/C++ Building Settings.**

## 10-4. Create a zynq_fsbl application.

**10-4-1.** Select **File** > **New** > **Application Project.**

**10-4-2.** Enter **zynq_fsbl** as the *Project Name*, and for *Board Support Package*, choose **Create New**.

**10-4-3.** Click **Next**, select *Zynq FSBL,* and click **Finish.**

   This will create the first stage bootloader application called zynq_fsbl.elf

## 10-5. Create a Zynq boot image.

**10-5-1.** Select **Xilinx Tools** > **Create Boot Image.**

**10-5-2.** Click the Browse button of the Output BIF file path field, browse to *c:\Summer\Tutorial,* and then click **Save** with the *output* as the default filename.

**10-5-3.** Click on the **Add** button of the *Boot image partitions,* click the Browse button in the Add Partition form, browse to *c:\Summer\Tutorial* **\tutorial\tutorial.sdk\zynq_fsbl\Debug** directory, select *zynq_fsbl.elf* and click **Open**.

**10-5-4.** Click **OK**. Click again on the **Add** button of the *Boot Image partitions,* click the Browse button in the Add Partition form, browse to *c:\Summer\Tutorial* **\Bitstreams** directory, select *BLANK.bit* and click **Open**.

**10-5-5.** Click **OK**.

**10-5-6.** Click again on the **Add** button of the *Boot Image partitions,* click the Browse button in the Add Partition form, browse to *c:\Summer\Tutorial* **\tutorial\tutorial.sdk\TestApp\Debug** directory, select *TestApp.elf* and click **Open**.

**10-5-7.** Click **OK**.

**10-5-8.** Make sure that the output path is *c:\Summer\Tutorial* and the filename is *BOOT.bin,* and click **Create Image**.

**10-5-9.** Close the SDK program by selecting **File > Exit**.

# Test the Design                                                                 Step 11

**11-1.** Connect the board with micro-USB cable connected to the UART. Place the board in the SD boot mode. Copy the generated BOOT.bin and the partial

bit files on the SD card and place the SD card in the board. Power On the board.

**11-1-1.** Make sure that a micro-usb cable is connected to the UART port.

**11-1-2.** Make sure that the board is set to boot in SD card boot mode.

**11-1-3.** Using the Windows Explorer, copy the **BOOT.bin** from the *c:\Summer\Tutorial* directory on to a SD Card. Also copy all the partial bitstreams into the SD card.

**11-1-4.** Place the SD Card in the board and power ON the board.

**11-2. Start a terminal emulator program such as TeraTerm or HyperTerminal. Select an appropriate COM port (you can find the correct COM number using the Control Panel). Set the COM port for 115200 baud rate communication.**

**11-2-1.** Start a terminal emulator program such as TeraTerm or HyperTerminal.

**11-2-2.** Select the appropriate COM port (you can find the correct COM number using the Control Panel).

**11-2-3.** Set the *COM* port for **115200** baud rate communication.

**11-2-4.** Reset the board and the menu will appear. Use it as you want.

```
    1: UCB TUNED
    2: UCB
    3: UCB VARIANCE
> 1
UCB Tuned Partial Binaries transferred successfully!
Blank Partial Binaries transferred successfully!
All Partial Binaries transferred successfully!
    1: Machine 1 Active
    2: Machine 2 Active
    3: Machine 3 Active
    4: Machine 4 Active
    5: Machine 3 Inactive
    6: Machine 4 Inactive
    7: Run the algorithm.
    0: Exit
> 1
Starting Machine 1 Reconfiguration
Machine 1 Reconfiguration Completed!
```

**Here, I have entered 1 to transfer the UCB_TUNED partial binaries from SD card to DDR3 memory. The success message is displayed.**

```
2
Starting Machine 2 Reconfiguration
Machine 2 Reconfiguration Completed!
    1: Machine 1 Active
    2: Machine 2 Active
    3: Machine 3 Active
    4: Machine 4 Active
    5: Machine 3 Inactive
    6: Machine 4 Inactive
    7: Run the algorithm.
    0: Exit
> 7
Machine 1 was selected 0 number of times and gave a reward of 0
Machine 2 was selected 5028 number of times and gave a reward of 10000
Machine 3 was selected 0 number of times and gave a reward of 0
Machine 4 was selected 0 number of times and gave a reward of 0
```

==Here, I presses 2 in order to activate machine 2 and then ran the algorithm. Since no other arm is active, the algorithm has selected machine 2 every time.==

```
    1: Machine 1 Active
    2: Machine 2 Active
    3: Machine 3 Active
    4: Machine 4 Active
    5: Machine 3 Inactive
    6: Machine 4 Inactive
    7: Run the algorithm.
    0: Exit
> 0
    1: UCB TUNED
    2: UCB
    3: UCB VARIANCE
> 3
UCB_V Partial Binaries transferred successfully!
Blank Partial Binaries transferred successfully!
All Partial Binaries transferred successfully!
    1: Machine 1 Active
    2: Machine 2 Active
    3: Machine 3 Active
    4: Machine 4 Active
    5: Machine 3 Inactive
    6: Machine 4 Inactive
    7: Run the algorithm.
    0: Exit
>
```

==When you're done, press 0 to Exit and then you can choose a different algorithm and the corresponding partial binaries will be transferred to the DDR to replace the old binaries.==

# Conclusion

This lab showed you steps involved in creating a processor system using Vivado IPI.  Full bitstream as well as partial reconfiguration bitstreams were generated by going through the PR flow.  You also learned how to generate the boot image as well as how to convert the partial bit files to bin format.  You verified the functionality using ZC706.

# Tutorial: Intelligent and Reconfigurable Architecture for Proposed KLUCB+UCB Algorithm

Authors: S. V. Sai Santosh and Dr. Sumit J. Darak, IIIT Delhi

{siripurapu17197, sumit}@iiitd.ac.in

## Introduction

In this lab, you will use Vivado IPI and Software Development Kit to create a reconfigurable peripheral using ARM Cortex-A9 processor system on Zynq. You will use Vivado IPI to create a top-level design, which includes the Zynq processor system as a sub-module. During the PR flow, you will define four Reconfigurable Partitions having three Reconfigurable Modules (KL_UCB, UCB and Comparator). You will create multiple Configurations and run the Partial Reconfiguration implementation flow to generate full and partial bitstreams. You will use ZC706 to verify the design in hardware using a SD card to initially configure the FPGA, and then partially reconfigure the device using the PCAP under user software control.

## Github Link: https://github.com/Sai-Santosh-99/PR_UCB

## Design Description

The purpose of this lab exercise is to implement a design that can be dynamically reconfigurable using PCAP resource and PS sub-system. The system consists of four peripheral (arms), having two unique function calculation capabilities (KL_UCB, UCB) and a reconfigurable Comparator design. The architecture is designed such that it switches automatically to a much resource-efficient UCB algorithm after learning the arm statistics using KL-UCB. The output (Quality-factor) of these machines goes into a Comparator IP which selects the machine with the largest Q-factor value. The user verifies the functionality using a user application. The dynamic modules are reconfigured using the PCAP resource available through Device Configuration block. The design is shown in Figure 1.

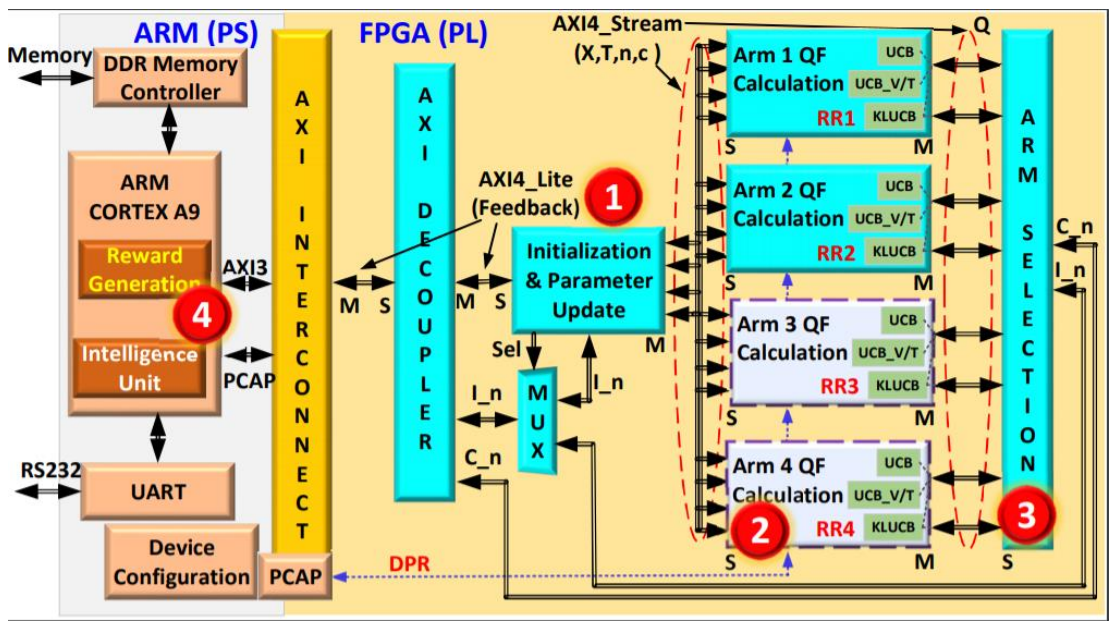

**Figure 1. The design**

The Sources directory provides the machine cores, source file for KL-UCB, UCB, the software application (C-code TestApp.c), and a place holder for the floorplan constraints (floorPlan.xdc). The Synth and its sub-directories structure will hold the synthesized checkpoints, the Implement and its sub-directories will hold the implemented configurations, the Checkpoint will hold the static, and the two configuration checkpoints, and the Bitstreams directory will hold the generated full and partial bitstreams. In the home directory, there are several TCL scripts which will perform several tasks including the processor system creation and the bottom-up synthesis of the reconfigurable modules.

## Procedure

This lab is separated into steps that consist of general overview statements that provide information on the detailed instructions that follow. Follow these detailed instructions to progress through the lab.

## General Flow for this Lab

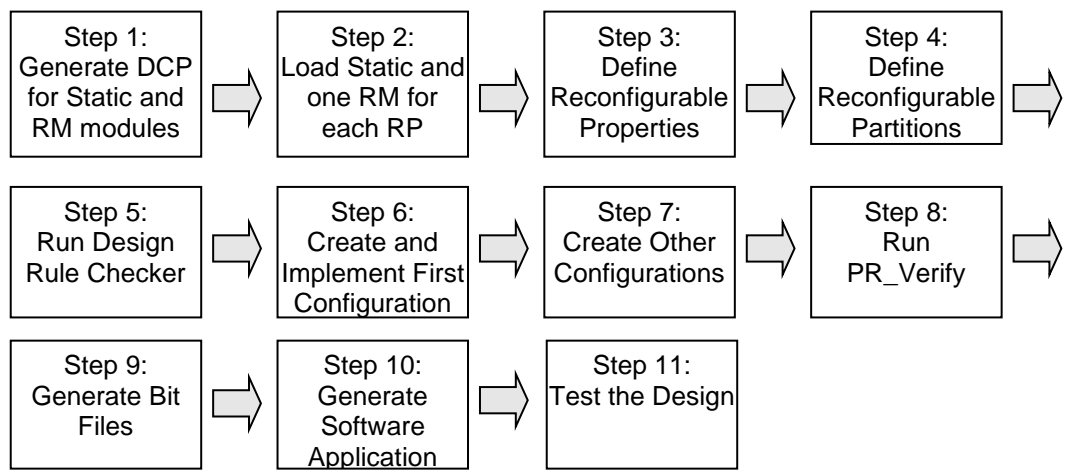

## Generate DCPs for the Static Design and RM Modules         Step 1

**1-1.    Start Vivado and execute the provided Tcl script to create the design check point for the static design having one RP.**

**1-1-1.** Open **Vivado** by selecting **Start > All Programs > Xilinx Design Tools > Vivado 2018.2 > Vivado 2018.2**

**1-1-2.** In the Tcl Shell window enter the following command to change to the lab directory and hit **Enter**.

```
cd c:/Summer/Tutorial
```

**1-1-3.** Generate the PS design executing the provided Tcl script.

```
source blockDesign.tcl
```

This script will create the block design called system, instantiate ZYNQ PS with SD 0 and UART 1 interfaces enabled. It will also enable the GP0 interface along with FCLK0 and RESET0_N ports. The provided machines IPs, PR decoupler and 2 Comparator IPs will also be instantiated. It will then create a top-level wrapper file called system_wrapper.v which instantiates the system.bd (the block design).

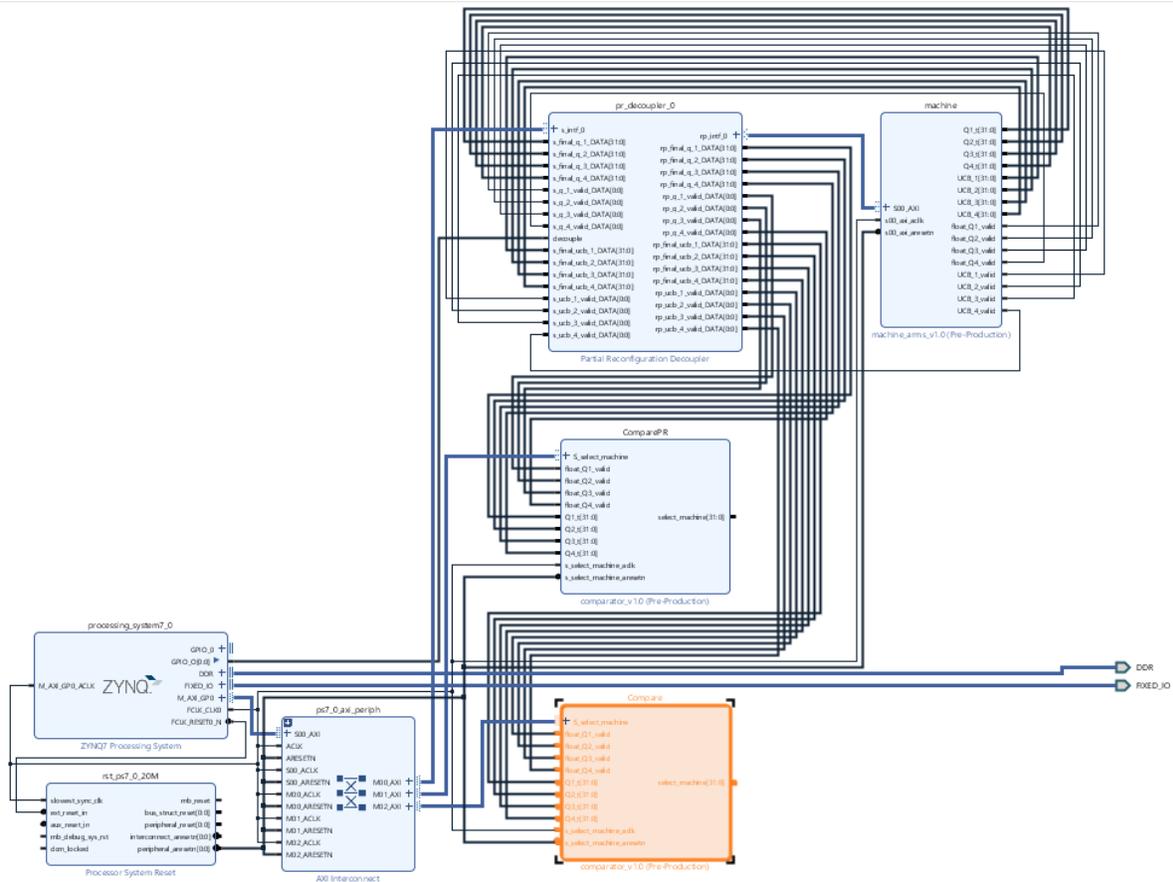

**Figure 2. The system block design**

**1-1-4.** Select the **Address Editor** tab. Expand **Data. Expand Unmapped Slaves, if any, and** right-click and select **Assign Address.**

**1-1-5.** Select **Tools** > **Validate Design.**

**1-1-6.** Select **File** > **Save Block Design.**

## 1-2. Synthesize the design to generate the dcp for the static logic of the design.

**1-2-1.** Click **Run Synthesis** under the *Synthesis* group in the *Flow Navigator* to run the synthesis process.

Wait for the synthesis to complete. When done click **Cancel**.

**1-2-2.** Using the windows explorer, copy the **system_wrapper.dcp** file from *tutorial\tutorial.runs\synth*_1 into the *Synth\Static* directory under the current lab directory.

**1-2-3.** Copy design checkpoints for the auto_pc, machine_arms, comparator_0, comparator_1, pr_decoupler_0, xbar_0, rst_ps7_0_20M, and processing_system7_0 instances to *Synth\Static* to sit alongside system_wrapper.dcp

**1-2-4.** Close the project by typing the `close_project` command in the Tcl console or selecting **File > Close Project**.

**1-3.** **Since we have RMs in HDL format, we need to synthesize them and generate the dcp for each of the RMs. The generated dcps should be stored in appropriate directories so they can be accessed correctly; particularly, the dcp files for RM must be in separate directories as their dcp file names will be same for a given RP.**

**1-3-1.** The HDL files for all the algorithms corresponding to each have been provided. Synthesize each each RM by creating a separate Vivado project. For the algorithm, synthesize the KL_UCB IP as kl_ucb, and synthesize the UCB IP as Q_function. You can also see these IPs instantiated in the given HDL files.

**1-3-2.** Synthesize each of the RMs and write the design checkpoint (dcp) in the respective destination folder under the Synth directory. After each RM's dcp is generated, close the design.

**1-3-3.** At this point the directory content will look like shown below. Here, UCB_1 denotes that this dcp corresponds to the UCB algorithm for machine 1. You just need to synthesize each algorithm separately and add the corresponding .dcp files to the respective folders.

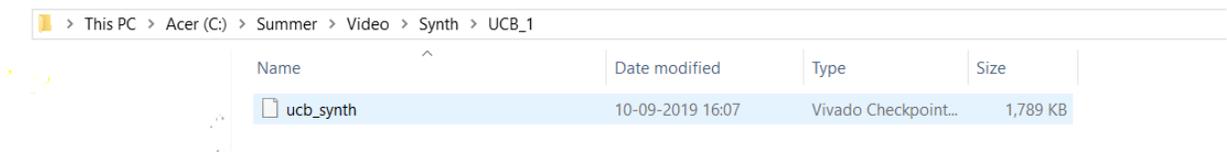

**Figure 7. Synth directory hierarchy and content**

**1-3-4.** Also, synthesize the dcp for both the comparator modules. One of the modules will be switched to blanking configuration once the architecture switches to the resource-efficient UCB algorithm.

## Load Static and one RM for the RP in Vivado                                Step 2

**Since all required netlist files (dcp) for the design are now available, you will use Vivado to floorplan the design, define Reconfigurable Partitions, add Reconfigurable Modules, run the implementation tools, and generate the full and partial bitstreams.**

**2-1.** **In this step you will load the static and one RM designs for the RP.**

**2-1-1.** In the Tcl Shell window enter the following command to change to the lab directory and hit **Enter**.

```
cd c:/Summer/Tutorial
```

**2-1-2.** Execute the following Tcl script to load the static design checkpoint.

```
source load_design_checkpoints.tcl
```

The script will do the following:

- Load the static design using the **open_checkpoint** command.

    ```
    open_checkpoint Synth/Static/system_wrapper.dcp
    ```

- Load the IP checkpoint for the Processing System by using the **read_checkpoint** command.

    ```
    read_checkpoint -cell system_i/processing_system7_0
    Synth/Static/system_processing_system7_0_0.dcp
    ```

- Load the IP checkpoint for the machines by using the **read_checkpoint** command.

    ```
    read_checkpoint -cell system_i/machine
    Synth/Static/system_machine_arms_0_0.dcp
    ```

- Load the IP checkpoint for the decoupler by using the **read_checkpoint** command.

    ```
    read_checkpoint -cell system_i/pr_decoupler_0
    Synth/Static/system_pr_decoupler_0_0.dcp
    ```

- Load the IP checkpoint for the Processing Reset by using the **read_checkpoint** command.

    ```
    read_checkpoint -cell system_i/rst_ps7_0_100M
    Synth/Static/system_rst_ps7_0_20M_0.dcp
    ```

- Load the IP checkpoint for the auto pc by using the **read_checkpoint** command.

    ```
    read_checkpoint -cell
    system_i/ps7_0_axi_periph/s00_couplers/auto_pc
    Synth/Static/system_auto_pc_0.dcp
    ```

- Load the IP checkpoint for the Comparator 1 by using the **read_checkpoint** command.

    ```
    read_checkpoint -cell system_i/ComparePR
    Synth/Static/system_comparatorPR_0_0.dcp
    ```

- Load the IP checkpoint for the Comparator 2 by using the **read_checkpoint** command.

    ```
    read_checkpoint -cell system_i/Compare
    Synth/Static/system_comparator_0_0.dcp
    ```

- Load the IP checkpoint for the system bus xbar by using the **read_checkpoint** command.

    ```
    read_checkpoint -cell system_i/ps7_0_axi_periph/xbar
    Synth/Static/system_xbar_0.dcp
    ```

**2-1-3.** Load one RM (KL-UCB) for the RP by using the **read_checkpoint** command.

```
read_checkpoint -cell
system_i/machine/inst/machine_arms_v1_0_S00_AXI_inst/u1
Synth/KL_1/ucb_synth.dcp

read_checkpoint -cell
system_i/machine/inst/machine_arms_v1_0_S00_AXI_inst/u2 Synth/
KL_2/ucb_synth.dcp

read_checkpoint -cell
system_i/machine/inst/machine_arms_v1_0_S00_AXI_inst/u3 Synth/
KL_3/ucb_synth.dcp
```

```
read_checkpoint -cell
system_i/machine/inst/machine_arms_v1_0_S00_AXI_inst/u4 Synth/
KL_4/ucb_synth.dcp

read_checkpoint -cell
system_i/ComparePR/inst/comparatorPR_v1_0_S00_AXI_inst/u1 Synth/
Compare/compare.dcp
```

You can now see the design structure in the Netlist pane with an RM for the *u1, u2, u3 and u4* module loaded.

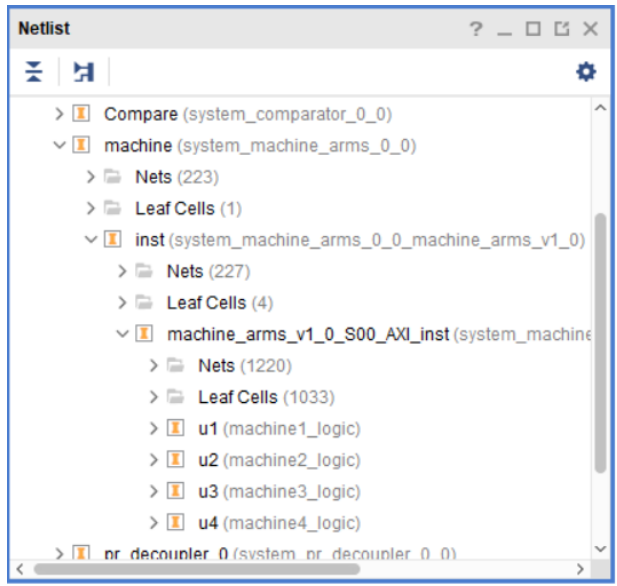

**Figure 9. Static design with RM loaded**

**2-1-4.** Select the *u1, u2 u3, u4 and u1(Comparator) instances one-by-one* and then select the *Properties* tab in the **Cell Properties** window. Note that the *IS_BLACKBOX* checkbox is not checked since a RM design is loaded.

## Define Reconfigurable Properties on each RM                         Step 3

**3-1.    In this design you have one Reconfigurable Partition having two RMs. Define the reconfigurable properties to the loaded RM.**

**3-1-1.** Define each of the loaded RMs (submodules) as partially reconfigurable by setting the **HD.RECONFIGURABLE** property using the following commands.

```
set_property HD.RECONFIGURABLE 1 [get_cells
system_i/machine/inst/machine_arms_v1_0_S00_AXI_inst/u1]

set_property HD.RECONFIGURABLE 1 [get_cells
system_i/machine/inst/machine_arms_v1_0_S00_AXI_inst/u2]

set_property HD.RECONFIGURABLE 1 [get_cells
system_i/machine/inst/machine_arms_v1_0_S00_AXI_inst/u3]

set_property HD.RECONFIGURABLE 1 [get_cells
system_i/machine/inst/machine_arms_v1_0_S00_AXI_inst/u4]
```

```
set_property HD.RECONFIGURABLE 1 [get_cells
system_i/ComparePR/inst/comparatorPR_v1_0_S00_AXI_inst/u1]
```

This is the point at which the Partial Reconfiguration license is checked.

**3-1-2.** Select the `u1, u2, u3, u4 and u1(Comparator)` instance and notice that the DONT_TOUCH checkbox is selected in the **Cell Properties** window.

## Define the Reconfigurable Partition Region

**3-2. Next you must floorplan the RP region.  Depending on the type and amount of resources used by all the RMs for the given RP, the RP region must be appropriately defined so it can accommodate any RM variant.**

**3-2-1.** You execute the following command to define the region for each RP, perform the DRC.

```
read_xdc floorplan.xdc
```

## Create and Implement First Configuration

**4-1. Create and implement the first Configuration.**

**4-1-1.** Execute the following command to implement the first configuration, the KL-UCB variant.

```
source create_first_configuration.tcl
```

The script will do the following tasks:

- The script will optimize, place and route the design by executing the following commands.

  ```
  opt_design
  
  place_design
  
  route_design
  ```

- Save the full design checkpoint.

  ```
  write_checkpoint -force Implement/KL/top_route_design.dcp
  ```

  At this point, a fully implemented partial reconfiguration design from which full and partial bitstreams can be generated is ready. The static portion of this configuration **must** be used for all subsequent configurations, and to isolate the static design, the current reconfigurable module must be removed.

**4-2. After the first configuration is created, the static logic implementation will be reused for the rest of the configurations.  So it should be saved. But before you save it, the loaded RM should be removed.**

**4-2-1.** Execute the following command to update the design with the blackbox and write the checkpoint.

```
source lock_placement_with_blackbox.tcl
```

The script will do the following tasks:

- Clear out the existing RMs executing the following commands.

  ```
  update_design -cell
  system_i/machine/inst/machine_arms_v1_0_S00_AXI_inst/u1 -
  black_box

  update_design -cell
  system_i/machine/inst/machine_arms_v1_0_S00_AXI_inst/u2 -
  black_box

  update_design -cell
  system_i/machine/inst/machine_arms_v1_0_S00_AXI_inst/u3 -
  black_box

  update_design -cell
  system_i/machine/inst/machine_arms_v1_0_S00_AXI_inst/u4 -
  black_box

  update_design -cell
  system_i/ComparePR/inst/comparatorPR_v1_0_S00_AXI_inst/u1 -
  black_box
  ```

  Issuing this command will result in design changes including, the number of Fully Routed nets (green) decreased, the number of Partially Routed nets (yellow) has increased, and *rp_instance* may appear in the Netlist view as empty.

- Lock down all placement and routing by executing the following command.

  ```
  lock_design -level routing
  ```

  Because no cell was identified in the `lock_design` command, the entire design in memory (currently consisting of the static design with black boxes) is affected.

- Write out the remaining static-only checkpoint by executing the following command.

  ```
  write_checkpoint -force Checkpoint/static_route_design.dcp
  ```

  This static-only checkpoint would be used for any future configuration, but here, you simply keep this design open in memory.

## Create Other Configurations

### 5-1. Read next set of RM dcp, create and implement the second configuration.

**5-1-1.** Execute the following command to create and implement the second configuration, the UCB_T variant.

```
source create_second_configuration.tcl
```

The script will do the following tasks:

- First, it will open the blanking configuration using the tcl command:

  ```
  open_checkpoint Checkpoint/static_route_design
  ```

- With the locked static design open in memory, read in post-synthesis checkpoint for the second reconfigurable module.

  ```
  read_checkpoint -cell
  system_i/machine1/inst/machine1_blank_v1_0_S00_AXI_inst/u1
  Synth/UCB_1/ucb_synth.dcp

  read_checkpoint -cell
  system_i/machine2/inst/machine2_blank_v1_0_S00_AXI_inst/u1
  Synth/UCB_2/ucb_synth.dcp

  read_checkpoint -cell
  system_i/machine3/inst/machine3_blank_v1_0_S00_AXI_inst/u1
  Synth/UCB_3/ucb_synth.dcp

  read_checkpoint -cell
  system_i/machine4/inst/machine4_blank_v1_0_S00_AXI_inst/u1
  Synth/UCB_4/ucb_synth.dcp

  read_checkpoint -cell
  system_i/ComparePR/inst/comparatorPR_v1_0_S00_AXI_inst/u1
  Synth/Compare/compare.dcp
  ```

  Optimize, place and route the design by executing the following commands.

  ```
  opt_design

  place_design

  route_design
  ```

- Save the full design checkpoint.

  ```
  write_checkpoint -force Implement/UCB/top_route_design.dcp
  ```

- Close the project

  ```
  close_project
  ```

## 5-2. Create the blanking configuration.

**5-2-1.** Execute the following command to create and implement the second configuration

```
source create_blanking_configuration.tcl
```

The script will do the following tasks:

- Open the static route checkpoint.

  ```
  open_checkpoint Checkpoint/static_route_design.dcp
  ```

- For creating the blanking configuration, use the `update_design -buffer_ports` command to insert LUTs tied to constants to ensure the outputs of the reconfigurable partition are not left floating.

  ```
  update_design -buffer_ports -cell
  system_i/machine/inst/machine_arms_v1_0_S00_AXI_inst/u1

  update_design -buffer_ports -cell
  system_i/machine/inst/machine_arms_v1_0_S00_AXI_inst/u2

  update_design -buffer_ports -cell
  system_i/machine/inst/machine_arms_v1_0_S00_AXI_inst/u3

  update_design -buffer_ports -cell
  system_i/machine/inst/machine_arms_v1_0_S00_AXI_inst/u4

  update_design -buffer_ports -cell
  system_i/ComparePR/inst/comparatorPR_v1_0_S00_AXI_inst/u1
  ```

- Now place and route the design. There is no need to optimize the design.

  ```
  place_design

  route_design
  ```

  The base (or blanking) configuration bitstream, when we generate in the next section, will have no logic for either reconfigurable partition, simply outputs driven by ground. Outputs can be tied to VCC if desired, using the HD.PARTPIN_TIEOFF property.

- Save the checkpoint in the `BLANK` directory.

  ```
  write_checkpoint -force Implement/BLANK/top_route_design.dcp
  ```

- Close the project

  ```
  Close_project
  ```

## Run PR_Verify

**6-1.** **You must ensure that the static implementation, including interfaces to reconfigurable regions, is consistent across all Configurations. To verify this, you run the PR_Verify utility**

**6-1-1.** Run the **pr_verify** command from the Tcl Console.

```
source verify_configurations.tcl
```

The script will perform the following tasks:

- Execute the pr_verify command and then close the project:

  ```
  pr_verify -initial Implement/KL/top_route_design.dcp -additional
  {Implement/BLANK/top_route_design.dcp
  Implement/UCB/top_route_design.dcp}
  ```

You should see the message indicating the KL configuration is compatible with BLANK, and the KL configuration is compatible with UCB. Execute the following command to close the project.

```
close_project
```

## Generate Bit Files

### 7-1. After all the Configurations have been validated by PR_Verify, full and partial bit files must be generated for the entire project

**7-1-1.** Generate the full configurations and partial bitstreams by executing the following tcl script.

```
source generate_bitstreams.tcl
```

**7-1-2.** The script will do the following tasks:

- Read the first configuration in the memory, using the command:

  ```
  open_checkpoint Implement/UCB/top_route_design.dcp
  ```

- Generate the full and partial bitstreams for this design.

  ```
  write_bitstream -file Bitstreams/UCB.bit

  write_cfgmem -format BIN -interface SMAPx32 -disablebitswap -
  loadbit "up 0 Bitstreams/UCB_pblock_machine1_partial.bit"
  Bitstreams/UCB_1.bin

  write_cfgmem -format BIN -interface SMAPx32 -disablebitswap -
  loadbit "up 0 Bitstreams/UCB_pblock_machine2_partial.bit"
  Bitstreams/UCB_2.bin

  write_cfgmem -format BIN -interface SMAPx32 -disablebitswap -
  loadbit "up 0 Bitstreams/UCB_pblock_machine3_partial.bit"
  Bitstreams/UCB_3.bin

  write_cfgmem -format BIN -interface SMAPx32 -disablebitswap -
  loadbit "up 0 Bitstreams/UCB_pblock_machine4_partial.bit"
  Bitstreams/UCB_4.bin

  write_cfgmem -format BIN -interface SMAPx32 -disablebitswap -
  loadbit "up 0 Bitstreams/UCB_pblock_Comparator_partial.bit"
  Bitstreams/Compare.bin

  close_project
  ```

  Notice the five bitstreams will be created.

**7-1-3.** Generate full and partial bitstreams for the second configuration.

- Open the checkpoint for KL-UCB using

  ```
  open_checkpoint Implement/KL/top_route_design.dcp

  write_cfgmem -format BIN -interface SMAPx32 -disablebitswap -loadbit "up 0 Bitstreams/ KL_pblock_machine1_partial.bit" Bitstreams/UCBV_1.bin

  write_cfgmem -format BIN -interface SMAPx32 -disablebitswap -loadbit "up 0 Bitstreams/ KL_pblock_machine2_partial.bit" Bitstreams/UCBV_2.bin

  write_cfgmem -format BIN -interface SMAPx32 -disablebitswap -loadbit "up 0 Bitstreams/ KL_pblock_machine3_partial.bit" Bitstreams/UCBV_3.bin

  write_cfgmem -format BIN -interface SMAPx32 -disablebitswap -loadbit "up 0 Bitstreams/ KL_pblock_machine4_partial.bit" Bitstreams/UCBV_4.bin

  close_project
  ```

  The five files will be created.

- Generate a full bitstream with black boxes, plus blanking bitstreams for the reconfigurable modules. Blanking bitstreams can be used to "erase" an existing configuration to reduce power consumption.

  ```
  write_bitstream -file Bitstreams/BLANK.bit

  write_cfgmem -format BIN -interface SMAPx32 -disablebitswap -loadbit "up 0 Bitstreams/BLANK_pblock_machine1_partial.bit" Bitstreams/BLANK_1.bin

  write_cfgmem -format BIN -interface SMAPx32 -disablebitswap -loadbit "up 0 Bitstreams/BLANK_pblock_machine2_partial.bit" Bitstreams/BLANK_2.bin

  write_cfgmem -format BIN -interface SMAPx32 -disablebitswap -loadbit "up 0 Bitstreams/BLANK_pblock_machine3_partial.bit" Bitstreams/BLANK_3.bin

  write_cfgmem -format BIN -interface SMAPx32 -disablebitswap -loadbit "up 0 Bitstreams/BLANK_pblock_machine4_partial.bit" Bitstreams/BLANK_4.bin

  write_cfgmem -format BIN -interface SMAPx32 -disablebitswap -loadbit "up 0 Bitstreams/BLANK_pblock_Comparator_partial.bit" Bitstreams/BLANK_Compare.bin

  close_project
  ```

- Generate full and partial bitstreams for the third configuration.

# About the C code

- This function is used for moving the partial binaries from the SD Card to the DDR memory.

```c
int SD_TransferPartial(char *FileName, u32 DestinationAddress, u32 ByteLength)
{
    FIL fil;
    FRESULT rc;
    UINT br;

    rc = f_open(&fil, FileName, FA_READ);
    if (rc) {
        xil_printf(" ERROR : f_open returned %d\r\n", rc);
        return XST_FAILURE;
    }

    rc = f_lseek(&fil, 0);
    if (rc) {
        xil_printf(" ERROR : f_lseek returned %d\r\n", rc);
        return XST_FAILURE;
    }

    rc = f_read(&fil, (void*) DestinationAddress, ByteLength, &br);
    if (rc) {
        xil_printf(" ERROR : f_read returned %d\r\n", rc);
        return XST_FAILURE;
    }

    rc = f_close(&fil);
    if (rc) {
        xil_printf(" ERROR : f_close returned %d\r\n", rc);
        return XST_FAILURE;
    }

    return XST_SUCCESS;
}
```

- This function is used to initialize the SD controller.

```c
int SD_Init()
{
    FRESULT rc;

    rc = f_mount(&fatfs, "", 0);
    if (rc) {
        xil_printf(" ERROR : f_mount returned %d\r\n", rc);
        return XST_FAILURE;
    }

    return XST_SUCCESS;
}
```

- Below, we have defined the function to write the data to our arms.

```
#define mWriteReg(BaseAddress, RegOffset, Data) \
    Xil_Out32((BaseAddress) + (RegOffset), (u32)(Data))
```

- Below, we have defined the function to read the data from the comparator.

```
#define mReadReg(BaseAddress, RegOffset) \
    Xil_In32((BaseAddress) + (RegOffset))
```

- Below is the pseudocode for the Multi-Armed Bandit algorithm on FPGA.

```
algo()
{
    for n times {

        •   Generate a random data value.

        •   Read the output from the Comparator

        •   If Comparator says Machine i is better-

            1. CHECK THE REWARD.

            2. INCREMENT X(i) if REWARD=1, else don't.

            3. INCREMENT T(i) as the Machine i was selected.

            4. WRITE THE CORRESPONDINT DATA TO THE FPGA CONVEYING THE
               MACHINE SELECTED AND ITS REWARD.

    }

}
```

## Generate the Software Application                                Step 10

**8-1.  Open the PS design that was created in Step 1. Export the hardware design and launch SDK.**

**8-1-1.**  Click on the **Open Project** link, browse to *c:/Summer/Tutorial/tutorial*, select the *tutorial.xpr* and click **OK** to open the design created in Step 1.

**8-1-2.**  Select **File > Export > Export Hardware…**

**8-1-3.**  In the *Export Hardware* form, <u>do not check</u> the *Include bitstream* checkbox and click **OK**.

**8-1-4.** Select **File > Launch SDK**

**8-1-5.** Click **OK** to launch SDK.

The SDK program will open. Close the Welcome tab if it opens.

## 8-2. Create a Board Support Package enabling generic FAT file system library.

**8-2-1.** In **SDK**, select **File** > **New** > **Board Support Package.**

**8-2-2.** Click **Finish** with the default settings (with standalone operating system).

This will open the Software Platform Settings form showing the OS and libraries selections.

**8-2-3.** Select **xilffs** as the FAT file support is necessary to read the partial bit files.

| Name | Version | Description |
|---|---|---|
| ☐ libmetal | 1.0 | Libmetal Library |
| ☐ lwip141 | 1.6 | lwIP TCP/IP Stack library: lwIP v1.4.1 |
| ☐ openamp | 1.1 | OpenAmp Library |
| ☑ xilffs | 3.4 | Generic Fat File System Library |
| ☐ xilflash | 4.2 | Xilinx Flash library for Intel/AMD CFI com… |
| ☐ xilisf | 5.7 | Xilinx In-system and Serial Flash Library |
| ☐ xilmfs | 2.1 | Xilinx Memory File System |
| ☐ xilpm | 2.0 | Power Management API Library for Zynq… |
| ☐ xilrsa | 1.2 | Xilinx RSA Library |
| ☐ xilskey | 6.0 | Xilinx Secure Key Library |

**Figure 13. Selecting the xilffs library support**

**8-2-4.** Click **OK** to accept the settings and create the BSP.

## 8-3. Create an application.

**8-3-1.** Select **File** > **New** > **Application Project.**

**8-3-2.** Enter **TestApp** as the *Project Name*, and for *Board Support Package*, choose **Use Existing** (*standalone_bsp_0* should be the only option).

**8-3-3.** Click **Next**, and select *Empty Application* and click **Finish.**

**8-3-4.** Expand the **TestApp** entry in the project view, right-click the *src* folder, and select **Import.**

**8-3-5.** Expand **General** category and double-click on **File System.**

**8-3-6.** Browse to *c:\Summer\Tutorial\Sources* and click **OK.**

**8-3-7.** Select **TestApp.c** and click **Finish** to add the file to the project.

**8-3-8.** Right-click on **TestApp** and select **C/C++ Building Settings.**

## 8-4. Create a zynq_fsbl application.

**8-4-1.** Select **File** > **New** > **Application Project.**

**8-4-2.** Enter **zynq_fsbl** as the *Project Name*, and for *Board Support Package*, choose **Create New**.

**8-4-3.** Click **Next**, select *Zynq FSBL,* and click **Finish.**

This will create the first stage bootloader application called zynq_fsbl.elf

## 8-5. Create a Zynq boot image.

**8-5-1.** Select **Xilinx Tools** > **Create Boot Image.**

**8-5-2.** Click the Browse button of the Output BIF file path field, browse to *c:\Summer\Tutorial,* and then click **Save** with the *output* as the default filename.

**8-5-3.** Click on the **Add** button of the *Boot image partitions,* click the Browse button in the Add Partition form, browse to *c:\Summer\Tutorial \tutorial\tutorial.sdk\zynq_fsbl\Debug* directory, select *zynq_fsbl.elf* and click **Open**.

**8-5-4.** Click **OK**. Click again on the **Add** button of the *Boot Image partitions,* click the Browse button in the Add Partition form, browse to *c:\Summer\Tutorial \Bitstreams* directory, select *BLANK.bit* and click **Open**.

**8-5-5.** Click **OK**.

**8-5-6.** Click again on the **Add** button of the *Boot Image partitions,* click the Browse button in the Add Partition form, browse to *c:\Summer\Tutorial \tutorial\tutorial.sdk\TestApp\Debug* directory, select *TestApp.elf* and click **Open**.

**8-5-7.** Click **OK**.

**8-5-8.** Make sure that the output path is *c:\Summer\Tutorial* and the filename is *BOOT.bin,* and click **Create Image**.

**8-5-9.** Close the SDK program by selecting **File > Exit**.

# Test the Design                                                                                                Step 11

**9-1.** **Connect the board with micro-USB cable connected to the UART. Place the board in the SD boot mode. Copy the generated BOOT.bin and the partial**

**bit files on the SD card and place the SD card in the board. Power On the board.**

**9-1-1.** Make sure that a micro-usb cable is connected to the UART port.

**9-1-2.** Make sure that the board is set to boot in SD card boot mode.

**9-1-3.** Using the Windows Explorer, copy the **BOOT.bin** from the *c:\Summer\Tutorial* directory on to a SD Card. Also copy all the partial bitstreams into the SD card.

**9-1-4.** Place the SD Card in the board and power ON the board.

**9-2. Start a terminal emulator program such as TeraTerm or HyperTerminal. Select an appropriate COM port (you can find the correct COM number using the Control Panel). Set the COM port for 115200 baud rate communication.**

**9-2-1.** Start a terminal emulator program such as TeraTerm or HyperTerminal.

**9-2-2.** Select the appropriate COM port (you can find the correct COM number using the Control Panel).

**9-2-3.** Set the *COM* port for **115200** baud rate communication.

**9-2-4.** Reset the board and the menu will appear. Use it as you want.

```
No. of enabled machines: 3
1. Enable a new arm.
2. Disable an arm.
3. An experiment for 10000 slots.
Enter the choice: 3

Starting new experiment with mean arm probabilities of 0.51, 0.52, 0.53
Architecture is switching from KL-UCB to UCB at 1526th slot.
Final result = Arms 1, 2, 3 were selected 2320, 4017, 3663 times.
****************************
1. Enable a new arm.
2. Disable an arm.
3. An experiment for 10000 slots.
Enter the choice: 1

Enabling arm.
No. of enabled arms: 4
****************************
1. Enable a new arm.
2. Disable an arm.
3. An experiment for 10000 slots.
Enter the choice: 3

Starting new experiment with mean arm probabilities of 0.51, 0.52, 0.53, 0.54
Architecture is switching from KL-UCB to UCB at 1809th slot.
Final result = Arms 1, 2, 3, 4 were selected 1523, 2810, 2421, 3246 times.
****************************
1. Enable a new arm.
2. Disable an arm.
3. An experiment for 10000 slots.
Enter the choice: 2

Disabling arm.
No. of enabled arms: 3
****************************
1. Enable a new arm.
2. Disable an arm.
3. An experiment for 10000 slots.
```

The snapshot of the terminal with example user experiments

## Conclusion

This lab showed you steps involved in creating a processor system using Vivado IPI. Full bitstream as well as partial reconfiguration bitstreams were generated by going through the PR flow. You also learned how to generate the boot image as well as how to convert the partial bit files to bin format. You verified the functionality using ZC706.

	
\end{document}